\documentclass[prl,twocolumn,nopacs,amsmath,amssymb,superscriptaddress]{revtex4-2}

\usepackage{amsmath}
\usepackage{amssymb}
\usepackage{amsbsy}
\usepackage{graphicx}
\usepackage{xcolor}

\usepackage{mathrsfs}

\usepackage{enumerate}
\usepackage{mathtools}

\usepackage{dcolumn}

\newcommand{\moy}[1]{\left\langle #1 \right\rangle}
\newcommand{\moyst}[1]{\left\langle #1 \right\rangle_{\mathrm{st}}}
\newcommand{\dd}[0]{\mathrm{d}}

\def\e{e}
\def\I{\mathrm{i}}

\newcommand{\dep}[2]{\ensuremath{\frac{\partial #1}{\partial #2}}}
\newcommand{\depn}[3]{\ensuremath{\frac{\partial^{#3} #1}{\partial #2^{#3}}}}

\usepackage{dsfont}
\def\un{\mathds{1}}
\def\lloc{\ell^{(\mathrm{loc})}}

\def\Z{\mathcal{Z}}
\def\pst{p_{\mathrm{st}}}
\def\Tr{\mathrm{Tr}}
\def\Div{\mathrm{div}}

\newcommand{\dmat}[1]{\delta^{(#1)}}

\definecolor{darkblue}{rgb}{0,0,0.6}
\definecolor{darkred}{rgb}{0.6,0,0}

\usepackage[colorlinks=true,urlcolor=darkblue,citecolor=darkblue,linkcolor=darkred]{hyperref}

\begin{document}

\title{Exact Lyapunov spectrum and localization length of the multichannel disordered Schrödinger equation}

\author{Aur\'elien Grabsch}
\affiliation{Sorbonne Universit\'e, CNRS, Laboratoire de Physique Th\'eorique de la Mati\`ere Condens\'ee (LPTMC), 4 Place Jussieu, 75005 Paris, France}

\begin{abstract}
  We consider the multichannel Schrödinger equation, describing the evolution of $N$ transverse modes in a finite width wire, in the presence of a Gaussian white noise potential in the unitary symmetry class.  For arbitrary energy $E$ and disorder strength $T$, we compute exactly the set of Lyapunov exponents $\{ \gamma_n(E,T) \}_{1 \leq n \leq N}$ which define the decay rates of the wave functions. In the large energy regime, we recover that the Lyapunov exponents are equally spaced, as predicted by the so-called DMPK approach. However we show that this property breaks down at arbitrary energy, giving rise to a richer Lyapunov spectrum. We also uncover a direct relation between the localization length, given by the inverse of the smallest Lyapunov exponent, and the density of states. Our derivation relies on the analysis of matrix stochastic differential equation for a modified Riccati matrix.
\end{abstract}

\maketitle

\emph{Introduction.---} Electronic transport in disordered systems has been extensively studied since the pioneering work of Anderson~\cite{Anderson:1958}. In a perfect crystal, the eigenstates of the Hamiltonian correspond to extended wave functions, while the presence of disorder might induce a localization of the states, characterized by an exponential decay of envelopes of the wave functions in space. Dimensionality plays a crucial role in the phenomenology of the Anderson localization~\cite{Kramer:1993,Evers:2008}: in one dimension, all the states are localized for any disorder, while a transition between localized and extended states might occur in higher dimension. However, exact results are mostly limited to the one dimensional case, for which powerful analytical techniques have been developed~\cite{Luck:1992,Texier:1999,Texier:2010}. One of the first exact solutions concerns the Schrödinger equation with a Gaussian white noise potential, which has become a paradigmatic model of disordered systems following the seminal works of Frisch, Lloyd and Halperin~\cite{Frisch:1960,Halperin:1965,Herbert:1971,Thouless:1972}. Since then, the density of states and the localization length, which quantifies the extension of the wave functions and the transport properties~\cite{Antsygina:1981,Lifshits:1988}, have been computed for various one-dimensional models~\cite{Ovchinnikov:1977,Nieuwenhuizen:1983,Bouchaud:1990,Barnes:1990,Comtet:1995,Hagendorf:2008,Comtet:2010,Comtet:2011,Comtet:2013,Grabsch:2014}. See also Refs.~\cite{Dumaz:2020,Dumaz:2023,Dumaz:2024b,Dumaz:2024} for related mathematical works.

To go beyond the strictly one dimensional case, models of multichannel disordered wires have been introduced to study the transport properties in systems of width much smaller than their length, so that only a finite number $N$ of transverse modes can propagate at a given energy. These quasi-1D models have become central in the theory of disordered systems, both for analytical and numerical studies~\cite{Kramer:1993,Beenakker:1997,Evers:2008}. From an analytical point of view, they are often studied using the framework introduced by Dorokhov, Mello, Pereyra and Kumar (DMPK)~\cite{Dorokhov:1982,Dorokhov:1982b,Dorokhov:1983,Dorokhov:1988,Mello:1988,Beenakker:1997,Mello:2004}. Their approach focuses on the analysis of the scattering matrix which characterizes the transport properties of the wire (see also~\cite{Ossipov:2018}). This method has permitted, under a weak-scattering assumption, to characterize the localization properties in these systems. In particular, it predicts that the Lyapunov exponents, which characterize the decay rate of the envelopes of the wave functions, are equally spaced~\cite{Beenakker:1997}.

However, few models of multichannel disordered wires can be solved explicitly to provide the exact dependence of the physical quantities on the microscopic parameters of the model. 
For instance, the density of states has been obtained for multichannel wires with chiral symmetry~\cite{Brouwer:2000,Titov:2001,Grabsch:2016b,Grabsch:2018}. For the multichannel Schrödinger equation, the density of states has first been studied for large number of channels~\cite{Allez:2015,Fyodorov:2020,Fyodorov:2026a} and has only been determined very recently for any number of channels~\cite{Fyodorov:2026,Fyodorov:2026b}. Concerning the Lyapunov exponents, their sum is known to be related to the density of states by the Thouless
relation~\cite{Herbert:1971,Thouless:1972,Lacroix:1986,Molinari:2013}, but the individual exponents have been obtained only perturbatively for small disorder~\cite{Derrida:1987,Romer:2010,Ludwig:2013}. However, the exact expression of the individual Lyapunov exponents for arbitrary parameters, in particular of the smallest exponent whose inverse defines the localization length, have remained out of reach for decades.

In this Letter, we solve this problem by determining exactly all the Lyapunov exponents for the multichannel Schrödinger equation in the presence of a Gaussian white noise potential in the unitary symmetry class. This provides the multichannel counterpart to the solution of the one dimensional model obtained by Frisch, Lloyd and Halperin in the 60s~\cite{Frisch:1960,Halperin:1965,Thouless:1972}.
Importantly, we obtain the localization length for any number of channels, disorder strength and energy of the eigenstate. In particular, our results show that the equal spacing of the Lyapunov exponents predicted by the DMPK approach holds only in the large energy regimes and breaks down at arbitrary energy. The derivation relies on the analysis of a matrix Langevin equation for a modified Riccati variable and can be extended to other models of multichannel disordered wires.

\emph{Model and main results.---} We consider the mutlichannel Schrödinger equation
\begin{equation}
  \label{eq:DefSchrod}
  \mathcal{H} \psi = E \psi
  \:,
  \quad
  \mathcal{H} = -  \un_N \partial_x^2 + W(x)
  \:,
\end{equation}
where $\psi$ is a $N$-component vector satisfying Dirichlet boundary conditions $\psi(0) = \psi(L) = 0$, $\un_N$ is the identity matrix and $W$ is a Gaussian white noise hermitian matrix potential, taken from the distribution $\exp[- \frac{1}{4T} \int \Tr[W^\dagger(x) W(x)] \dd x]$, with $T = J^2/(2N)$ the amplitude of the noise. That is, $\moy{W(x)} = 0$ and $\moy{W_{ij}(x) W_{k,l}(x')} = 2 T \delta_{i,l} \delta_{j,k} \delta(x-x')$. We also introduce the $N \times N$ matrix $\Psi(x)$ solution of~(\ref{eq:DefSchrod}) on the positive axis, with the initial condition $\Psi(0) = 0$ and $\Psi'(0) = \un_N$. This matrix regroups the $N$ independent solutions of~(\ref{eq:DefSchrod}) satisfying the Dirichlet condition at the origin.

For $L \to \infty$, the wave functions are localized: their envelopes present an exponential decays at infinity, with rates $\{\gamma_n(E,T) \}_{1 \leq n \leq N}$ which are called Lyapunov exponents. These exponents can be defined from the fundamental solution $\Psi$ as the large $L$ behavior of the eigenvalues of the matrix $\Psi^\dagger(L) \Psi(L)$, more precisely~\cite{Oseledec:1968,Kaimanovich:1989}
\begin{equation}
  \label{eq:DefLyaps0}
  \text{e.v. of }
  \frac{1}{2L} \ln [\Psi^\dagger(L) \Psi(L)]
  \underset{L \to \infty}{\longrightarrow}
  \{\gamma_n(E,T) \}
  \:.
\end{equation}

In this article, we obtain an explicit expression for the Lyapunov exponents, for any number $N$ of channels, energy $E$ of the eigenmode and disorder strength $T$,
\begin{equation}
  \label{eq:Lyapn}
  \gamma_k(E,T) = \frac{T}{2} \partial_E \ln \frac{\Z_{N,k-1}(E,T)}{\Z_{N,k}(E,T)}
  \:,
\end{equation}
where we defined
\begin{equation}
  \label{eq:defZnk}
  \Z_{N,k}
  = \int_0^\infty \dd^k h
  \prod_{i<j} \frac{(h_i-h_j)^2}{h_i+h_j}
  \prod_{p=1}^k w_{N-k}(h_p)
  \:,
\end{equation}
with
\begin{equation}
  \label{eq:DefWeight}
  w_q(h) = h^{q-\frac{1}{2}} \e^{- \frac{1}{T}(\frac{h^3}{12} + E h)}
  \:.
\end{equation}

For multichannel disordered wires, it is well known that the sum of the Lyapunov exponents is related to the density of states $\rho_N(E,T)$ by the Thouless relation~\cite{Thouless:1972,Lacroix:1986,Molinari:2013}. This is however not sufficient to determine the localization length $\lloc(E,T)$, which is given by the inverse of the smallest Lyapunov
exponent. Here, we show that the smallest Lyapunov exponent is directly related to the integrated density of states per unit length $\nu_N(E,T) = \int_{-\infty}^E \rho_N(E',T) \dd E'$ by the strikingly simple relation
\begin{equation}
  \label{eq:RelLocDos}
  \frac{1}{\lloc(E,T)}
  \equiv
  \gamma_N(E,T)  = \frac{T}{2} \partial_E \ln \nu_N(E,T)
  \:.
\end{equation}
This relation gives a simple and direct way to compute and analyze the localization length from the recent results obtained on the density of states~\cite{Fyodorov:2026}.

\begin{figure}
  \centering
  \includegraphics[width=\columnwidth]{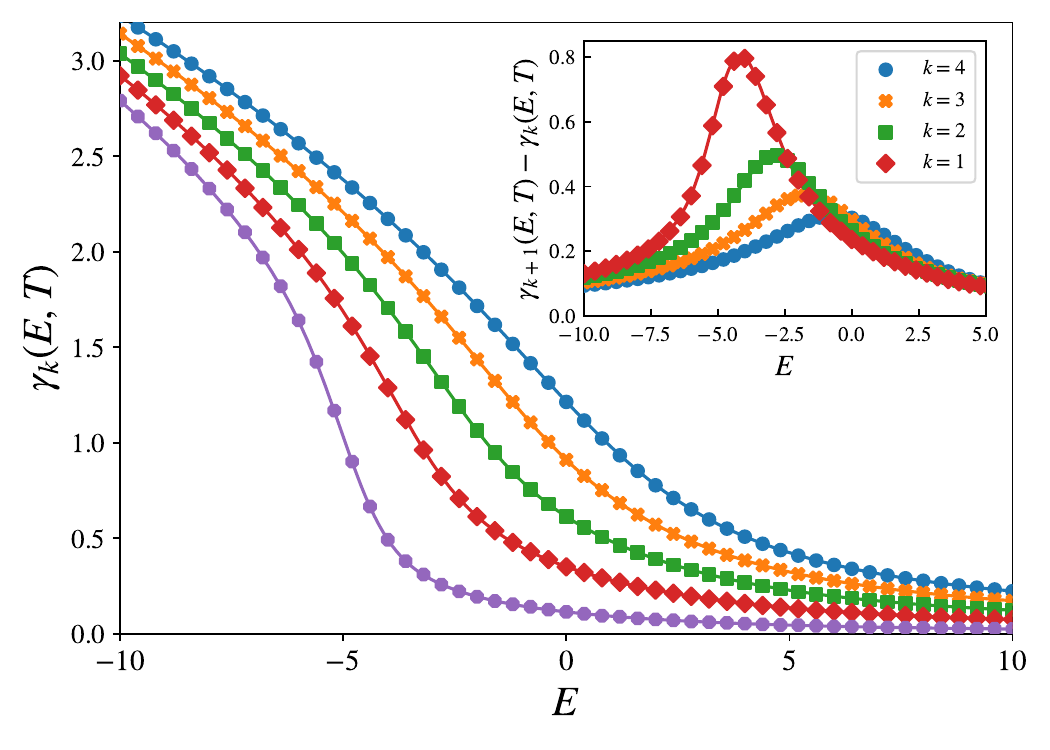}
  \caption{Lyapunov exponents for a disordered channel with $N=5$ modes with $T=1$, as a function of the energy. The localization length is given by the inverse of the smallest Lyapunov exponent. The points are obtained from numerical simulations detailed in the Supplemental Material~\cite{SM} and the lines from the analytical result~(\ref{eq:Lyapn}). The inset represents the spacings between consecutive Lyapunov exponents $\gamma_{k+1} - \gamma_k$. The spacings become equal for large $|E|$, but not for any $E$.}
  \label{Fig:Lyaps5}
\end{figure}

The Lyapunov exponents~(\ref{eq:Lyapn}) are represented in Fig.~\ref{Fig:Lyaps5} for $N=5$ and $T=1$ as a function of the energy $E$, and compared to numerical simulations. In the following, we first derive these results and then analyze them in more details.

\emph{Lyapunov exponents from a matrix stochastic process.---}
Following the ideas used for strictly one-dimensional channels ($N=1$)~\cite{Frisch:1960,Halperin:1965,Bouchaud:1990,Barnes:1990,Hagendorf:2008,Comtet:2010,Comtet:2011,Comtet:2013,Grabsch:2014,Luck:1992} and extended to multichannel wires~\cite{Allez:2015,Grabsch:2016b,Fyodorov:2020,Fyodorov:2026a,Fyodorov:2026,Fyodorov:2026b,Grabsch:2016b,Grabsch:2018}, it is convenient to introduce the Riccati matrix $Z = (\partial_x\Psi) \Psi^{-1}$, which obeys the stochastic equation
\begin{equation}
  \label{eq:Riccati}
  \partial_x Z = - Z^2 - E \un_N + W
  \:,
\end{equation}
with $Z(0) = \infty \un_N$. From this equation, we deduce that $Z^\dagger = Z$. The distribution of the noise matrix $W$ being invariant under unitary transformations, the dynamics of the eigenvalues and eigenvectors of $Z$ decouple and one can focus on the eigenvalues only. When one eigenvalue reaches $-\infty$ (due to the force $-Z^2$), it is reinjected at $+\infty$. The stationary distribution of the eigenvalues of $Z$ has been determined first in the large $N$ limit~\cite{Allez:2015,Fyodorov:2020,Fyodorov:2026a} and very recently for arbitrary number $N$ of channels~\cite{Fyodorov:2026}. From the distribution of $Z$, one can extract the density of states and the sum of the Lyapunov exponents~\cite{Grabsch:2016b,Grabsch:2018,Fyodorov:2026}, but not the individual exponents.

To compute each Lyapunov exponent individually, we rely on a QR decomposition of the fundamental solution $\Psi(x)$. The QR decomposition is a standard method for numerical computations of Lyapunov exponents~\cite{Dieci:1995,Dieci:2010}, but it can also be used for analytical calculations~\cite{Grabsch:2016b,Grabsch:2018}. We briefly sketch the main ideas behind this approach, before deriving the main equations.

Consider any initial vector $\psi_1(0)$, and construct $\psi_1(x)$ as the solution of $\partial_x \psi_1 = Z \psi_1$. Due to the evolution equation of  $Z$~\eqref{eq:Riccati}, this vector satisfies the Schrödinger equation~\eqref{eq:DefSchrod}. The norm of this vector will typically grow exponentially, with a rate given by the largest Lyapunov exponent $\gamma_1$, $|| \psi_1(x) || \sim \e^{\gamma_1 x}$. To obtain the second largest Lyapunov exponent $\gamma_2$, we need to follow two independent vectors $(\psi_1, \psi_2)$. Individually, each will grow exponentially with rate $\gamma_1$, but the volume they enclose will grow with rate $\gamma_1 + \gamma_2$. Extending this argument to $k$ independent vectors, taken to be the columns of $\Psi$, we get~\cite{Oseledec:1968,Kaimanovich:1989}
\begin{equation}
  \label{eq:AlterDefLyaps}
  \lim_{L \to \infty} \frac{1}{2L} \ln | \Psi^\dagger(L) \Psi(L) |_k
  = \sum_{n=1}^k \gamma_n
  \:,
\end{equation}
with $\gamma_1 > \gamma_2 > \cdots > \gamma_N$ and $| \Psi^\dagger \Psi |_k$ is the determinant of the top-left block of $\Psi^\dagger \Psi$ of size $k$, which represents the volume enclosed by the first $k$ column vectors. This gives an equivalent definition of the Lyapunov exponents to~(\ref{eq:DefLyaps0}). The volume $| \Psi^\dagger \Psi |_k$ can equivalently be computed by constructing an orthogonal basis from the first $k$ columns of $\Psi$, for instance using a Gram–Schmidt algorithm. This is exactly what a QR decomposition does. Explicitly, we decompose $\Psi$ as
\begin{equation}
  \label{eq:QRforPsi}
  \Psi(x) = Q(x) R(x)
  \:,
\end{equation}
where $Q$ is a unitary matrix and $R$ a upper triangular matrix with real diagonal elements. Differentiating this relation and using the definition of $Z$, we obtain $Q^\dagger \partial_x Q + (\partial_x R) R^{-1} = Q^\dagger Z Q$. Isolating the diagonal elements and lower triangular part of this equation, we obtain the evolution equations for $R$ and $Q$~\footnote{Since $Q$ is unitary, $Q^\dagger Q = \un_N$ and thus $Q^\dagger \partial_x Q =- (Q^\dagger \partial_x Q)^\dagger$, so that $\mathrm{Re}[(Q^\dagger \partial_x Q)_{nn}] = 0$.},
\begin{equation}
  \label{eq:QRevol}
  \partial_x \ln | R_{nn} | = (Q^\dagger Z Q)_{nn}
  \:,
  \quad
  Q^\dagger \partial_x Q = A_N \circ (Q^\dagger Z Q)
  \:,
\end{equation}
where $\circ$ denotes the entrywise Hadamard product $(A \circ B)_{ij} = A_{ij} B_{ij}$ and $A_N$ is $N \times N$ the antisymmetric matrix with elements $A_{ij} = 1$ if $i > j$, $A_{ij} = -1$ if $j>i$ and $A_{jj} = 0$. Since $| \Psi^\dagger \Psi |_k = |R^\dagger R|_k = \prod_{n=1}^k |R_{nn}|^2$ because $R$ is upper triangular, the equation for the Lyapunov exponents~(\ref{eq:AlterDefLyaps}) gives
\begin{equation}
  \label{eq:LyapsFromR}
  \gamma_k
  = \lim_{L \to \infty} \frac{1}{L} \ln |R_{kk}(L)|
  = \lim_{L \to \infty} \frac{1}{L} \int_0^L (Q^\dagger Z Q)_{kk}
  \:.
\end{equation}
The Lyapunov exponents are therefore determined by the joint dynamics of $Z$ and $Q$, and not by $Z$ only.

Since the Lyapunov exponents~(\ref{eq:LyapsFromR}) and the evolution equations~(\ref{eq:QRevol}) only involve $Q^\dagger Z Q$, we can simplify the problem by considering the matrix $Y \equiv Q^\dagger Z Q$, which following Itô's rule obeys 
\begin{equation}
  \label{eq:EvolY}
  \partial_x Y = - Y^2 - E \un_N
  + [Y, A_N \circ Y] + W
  \:.
\end{equation}
Here we have replaced $Q^\dagger W Q$ by $W$ since these two matrices have the same law and are uncorrelated with $Y$ in the Itô convention. Additionally, assuming that $Y$ admits a stationary distribution and denoting $\moy{\cdots}_{\mathrm{st}}$ the average with respect to this distribution, the Lyapunov exponents~(\ref{eq:LyapsFromR}) follow from
\begin{equation}
  \label{eq:LyapsFromY}
  \gamma_k(E,T) = \moy{Y_{kk}}_{\mathrm{st}}
  \:.
\end{equation}
Together, Eqs.~(\ref{eq:EvolY}) and~(\ref{eq:LyapsFromY}) fully determine the Lyapunov spectrum.  The matrix $Y$ is a modified version of the Riccati matrix $Z$, rotated by the unitary matrix $Q$. This additional rotation is responsible for the commutator term involved in the equation for $Y$~(\ref{eq:EvolY}) which does not appear in the equation for $Z$~(\ref{eq:Riccati}). This leads to a coupling between the eigenvalues and eigenvectors of $Y$ and makes it a more complicated object to study than the original Riccati matrix $Z$. It is however the good object to obtain the Lyapunov exponents.

\emph{Stationary distributions and explicit results.---} To compute the Lyapunov exponents from~\eqref{eq:LyapsFromY}, there remains to obtain the stationary distribution of $Y$. This is a difficult task, which we have not been able to perform. However, it is still possible to obtain the Lyapunov exponents by performing an additional change of basis. We decompose $Y$ into a block structure
\begin{equation}
  Y =
  \begin{pmatrix}
    Y_1 & Y_2
    \\
    Y_2^\dagger &  Y_3
  \end{pmatrix}
  \:,
  \quad
  Y_1 \text{ of size } k \times k
  \:.
\end{equation}
We introduce the unitary matrix $U = \mathrm{Diag}(U_1, U_2)$, with $U_1 \in U(k)$ and $U_2 \in U(N-k)$ defined by $U_1^\dagger \partial_x U_1 = A_{k} \circ Y_1$, $U_2^\dagger \partial_x U_2 = A_{N-k} \circ Y_3$. We also introduce the matrix
\begin{equation}
  \label{eq:MatriX}
  X = U Y U^\dagger = 
  \begin{pmatrix}
    X_1 & X_2
    \\
    X_2^\dagger &  X_3
  \end{pmatrix}
=
\begin{pmatrix}
    U_1 X_1 U_1^\dagger & U_1 X_2 U_2^\dagger
    \\
    U_2 X_2^\dagger U_1^\dagger &  U_2 X_3 U_2^\dagger
  \end{pmatrix}
  \:.
\end{equation}
This unitary transformation essentially removes, within each diagonal subblock, the effect of the rotation matrix $Q$. This leads to simpler equations satisfied by $X_1$ and $X_2$, which decouple from $X_3$,
\begin{subequations}
  \label{eq:EvolX}
  \begin{align}
    \label{eq:EvolX1}
    \partial_x X_1
    &= -X_1^2 + X_2 X_2^\dagger - E \un_k + W_1
      \:,
    \\
    \label{eq:EvolX2}
    \partial_x X_2
    &=
      -2 X_1 X_2 + W_2
      \:,
  \end{align}
\end{subequations}
where we have denoted $W_i$ the blocks of $W$ as in~(\ref{eq:MatriX}). The change of basis still allows us to determine the Lyapunov exponents, since from~(\ref{eq:LyapsFromY}),
\begin{equation}
  \label{eq:LyapsFromX}
  \sum_{n=1}^k \gamma_n(E,T) = \moyst{\Tr[Y_1]} = \moyst{\Tr[X_1]}
  \:.
\end{equation}

There remains to determine the stationary distribution $\pst(X_1,X_2)$. It is solution of the Fokker-Planck equation $\mathcal{L}^\dagger \pst = 0$, where the operator $\mathcal{L}^\dagger$ can be written by isolating the independent real components of the stochastic equations~\eqref{eq:EvolX} (see Supplemental Material (SM)~\cite{SM} for details),
\begin{multline}
  \label{eq:FokkerPlanck}
  \mathcal{L}^\dagger \pst =
  T \Delta_{X_1} \pst + \frac{T}{2} \Delta_{X_2} \pst
  + \Div_{X_2}[ 2 X_1 X_2 \pst]
  \\
  + \Div_{X_1}[ (X_1^2 - X_2 X_2^\dagger + E \un_k ) \pst ]
  \:,
\end{multline}
where the divergence and Laplace operators are defined explicitly in the End Matter below. We found that the solution takes the form (see also End Matter)
\begin{multline}
  \label{eq:JointStatioDistr}
  \pst(X_1,X_2) = \frac{1}{\mathcal{N}}
  \int_{H > 0} \dd H \:
  (\det H)^{2(N-k)}
  \\
  \times \e^{- \frac{1}{T} \Tr[H X_1^2 - X_1 H^2 + \frac{H^3}{3} + E H + H X_2 X_2^\dagger]}
  \:,
\end{multline}
where the integral runs over the $k \times k$ Hermitian matrix $H$ with positive eigenvalues and $\mathcal{N}$ is a normalization constant. For $k=N$, Eq.~\eqref{eq:JointStatioDistr} reduces to the stationary distribution of the Riccati matrix $Z$~(\ref{eq:Riccati}) found in~\cite{Fyodorov:2026} (written in Ref.~\cite{Fyodorov:2026b} in a form similar to~(\ref{eq:JointStatioDistr})).

From~(\ref{eq:LyapsFromX}), the Lyapunov exponents are obtained by averaging $\Tr[X_1]$ with the distribution~(\ref{eq:JointStatioDistr}). 
This computation can be performed explicitly, since the integrals over $X_1$ and $X_2$ are Gaussian. There remains only the integral over the eigenvalues of $H$, which gives (see SM~\cite{SM})
\begin{equation}
  \moyst{\Tr[X_1]} = - \frac{T}{2} \frac{\partial_E \Z_{N,k}(E,T)}{\Z_{N,k}(E,T)}
  \:,
\end{equation}
with $\Z_{N,k}$ defined in Eq.~(\ref{eq:defZnk}). From Eq.~(\ref{eq:LyapsFromX}), this gives the sum of the first $k$ Lyapunov exponents. Taking the difference between the expression for $k$ and $k-1$ gives the $k^{\mathrm{th}}$ Lyapunov exponent~(\ref{eq:Lyapn}) announced above. For $N=k=1$, we recover the known expression of the Lyapunov exponent for the one-dimensional Schrödinger equation with a white noise potential~\cite{Lifshits:1988,Texier:1999,Dumaz:2024b}.

\emph{Relation with the density of states.---} The integrated density of states per unit length $\nu_N(E,T)$ has recently been computed in~\cite{Fyodorov:2026} from the stationary distribution of the Riccati matrix $Z$, with extensions to other symmetry classes derived in~\cite{Fyodorov:2026b}. The general idea is that the density of states can be obtained from the fundamental solution $\Psi$ using a node-counting argument: the vanishing of $\det \Psi(L)$ indicates that a linear combination $\psi$ of the columns of $\Psi$ satisfies the second Dirichlet condition $\psi(L) = 0$. As a consequence, counting the divergences of the eigenvalues of the Riccati matrix $Z$ gives the number of eigenstates below $E$. Following this analysis, the integrated density of states $\nu_N$ is expressed in~\cite{Fyodorov:2026b} as (see also SM~\cite{SM} for a derivation from the stationary distribution~\eqref{eq:JointStatioDistr})
\begin{equation}
  \nu_N(E,T) = N \sqrt{\frac{T}{\pi}} \frac{\Z_{N,N-1}(E,T)}{\Z_{N,N}(E,T)}
  \:.
\end{equation}
Comparing with the expression of the Lyapunov exponents~(\ref{eq:Lyapn}), we obtain relation~(\ref{eq:RelLocDos}) between the smallest Lyapunov exponent and the density of states.

This is a strikingly simple relation between two quantities that are not expected to be so directly related. The density of states is well-known to be related to the sum of the Lyapunov exponents via the Thouless relation~\cite{Thouless:1972,Lacroix:1986,Molinari:2013}, but this does not determine the localization length. Here, we find that the localization length is directly related to the density of states by~(\ref{eq:RelLocDos}). This constitutes an important simplification, since the density of states can be extracted from the Riccati matrix~(\ref{eq:Riccati}), while the individual Lyapunov exponents can only be obtained from the more complex matrix $Y$~(\ref{eq:EvolY}). We use this relation below to analyze the behavior of the localization length for large number of channels.

\emph{Asymptotic behaviors of the Lyapunov exponents.---} The
asymptotic behavior of the Lyapunov exponents can be obtained from the
analysis of the integrals $\Z_{N,k}$~(\ref{eq:defZnk}). The details of the computations are given in the SM~\cite{SM}.
For fixed $N$ and $E \to -\infty$, we find $\Z_{N,k} \sim (-E)^{\frac{N k}{2} - k^2} \e^{\frac{4k}{3T}(-E)^{3/2}}$, so that the Lyapunov exponents~(\ref{eq:Lyapn}) read
\begin{equation}
  \label{eq:LyapsLargeNegE}
  \gamma_k(E,T) \underset{E \to -\infty}{\simeq}
  \sqrt{-E}
  + \frac{T}{4(-E)}(N - 4k + 2)
  \:.
\end{equation}
At large negative energy, the leading behavior is the disorderless contribution $\sqrt{-E}$ and the first correction from the disorder give an equal spacing of the Lyapunov exponent. At high energy $E \to \infty$, we find $\Z_{N,k} \sim E^{-N k + \frac{k^2}{2}}$, so that~(\ref{eq:Lyapn}) yield
\begin{equation}
  \label{eq:LyapsHighE}
  \gamma_k(E,T) \underset{E \to +\infty}{\simeq}
  \frac{T}{2E}\left( N - k + \frac{1}{2} \right)
  \:,
\end{equation}
with again equally spaced Lyapunov exponents. We recover with~\eqref{eq:LyapsHighE} the prediction of the DMPK approach~\cite{Beenakker:1997}, but only in the high energy regime. For arbitrary parameters, the Lyapunov exponents are not equally spaced, as shown in the inset of Fig.~\ref{Fig:Lyaps5}.

For large number of channels, with $J^2 = 2 N T$ fixed, the behavior of the localization length can be obtained from the results on the density of states~\cite{Allez:2015,Fyodorov:2020,Fyodorov:2026a,Fyodorov:2026} thanks to the relation~(\ref{eq:RelLocDos}). Denoting $s = (2J^2)^{1/3}$ and $E_\star = -3s^2/4$, we find that the smallest Lyapunov exponents, whose inverse defines the localization length, behaves as
\begin{equation}
  \gamma_N\left(E, \frac{J^2}{2N} \right)
  \underset{N \to \infty}{\simeq}
  \left\lbrace
    \begin{array}{ll}
      \sqrt{d(E)^2 - \frac{J^2}{4 d(E)}}
      & \text{for } E < E_\star \:,
      \\
      0 & \text{for } E > E_\star \:,
    \end{array}
  \right.
\end{equation}
where $d(E)$ is the largest real solution of the cubic equation $d^3 + E d + J^2/2 = 0$.
The crossover between these two regions is given by the scaling function
\begin{equation}
  \label{eq:LocLengthLargeN}
  N^{1/5} \gamma_N\left(E_\star + \frac{s^2 x}{N^{4/5}}, \frac{J^2}{2N} \right)
  \underset{N \to \infty}{\simeq} \frac{1}{4} \left( \frac{J}{2} \right)^{^\frac{2}{3}}
  \frac{\Phi'(x)}{\Phi(x)}
  \:,
\end{equation}
where $\Phi(x)$ is obtained from a specific solution of the Painlevé I equation~\cite{Fyodorov:2026}, whose definition is recalled in the End Matter. 
The $N^{1/5}$ scaling was argued in Ref.~\cite{Fyodorov:2026} from a Thouless  argument. Our relation~\eqref{eq:RelLocDos} confirms this scaling and provides the full crossover profile of the smallest Lyapunov exponent.

The vanishing of the smallest Lyapunov exponent at the critical energy $E_\star$ does not indicate a delocalization transition because the density of states per channel $\nu_N(E,T)/N$ also vanishes for
$E < E_\star$ so that there are effectively no states in this domain~\cite{Allez:2015,Fyodorov:2020,Fyodorov:2026a,Fyodorov:2026}. This is similar to the disorderless case $T=0$ in which $\gamma_k = \sqrt{-E}$ for $E < 0$ and vanishes for $E>0$, with the integrated density of states $\nu_N(E) = N \sqrt{E}/\pi$ for $E>0$. Note however that here the smallest Lyapunov exponent vanishes with a different power, due to the disorder, as $\lim_{N \to \infty} \gamma_N(E,J^2/(2N)) \sim (E_\star - E)^{1/4}$ for $E \to E_\star$.

\emph{Discussion and conclusion.---} We have obtained the exact Lyapunov spectrum for the multichannel Schrödinger equation with a random matrix potential in the unitary symmetry class (corresponding to the Dyson index $\beta=2$). Our analysis relies on the study of the stochastic matrix process~(\ref{eq:EvolY}) which, unlike the usual Riccati process~(\ref{eq:Riccati}) considered in the literature~\cite{Allez:2015,Grabsch:2016b,Fyodorov:2020,Fyodorov:2026a,Fyodorov:2026,Fyodorov:2026b}, contains the information on the localization properties. This allows us to go beyond the standard DMPK approach and obtain the dependence of the localization length on the parameters of the model. In particular, we have showed that the equal spacing of the Lyapunov exponents predicted by DMPK holds only in the high energy regime, with a richer structure at arbitrary energy.

Although the model is not strictly one-dimensional, it does not present an Anderson transition between localized and extended states: for any number of channel and disorder strength, the localization length remains finite. This is expected due to the fact that the matrix potential $W$ is taken to be isotropic, so that it couples all the channels equivalently (as in the DMPK approach), and thus describes a quasi-1D wire. Our formalism can be applied to study disordered wires with anisotropic channels, but obtaining a full solution remains a challenging task.open

Finally, we have restricted our analysis to the unitary symmetry class ($\beta=2$) for simplicity, but extensions to orthogonal ($\beta=1$) and symplectic ($\beta=4$) are expected, in particular since the
density of states for all these classes has recently been determined~\cite{Fyodorov:2026b}.

\emph{Acknowledgements.---} I am grateful to Yan Fyodorov for sharing the draft of Ref.~\cite{Fyodorov:2026b} and for stimulating exchanges. 
I acknowledge support from the European Research Council (ERC) Synergy Grant No. 101224739 (UniCIPS). 
I made use of ChatGPT (OpenAI) for preliminary investigations concerning the solution of the Fokker-Planck equation. All the calculations and analysis reported in this article, as well as the writing, has been performed by the author.


\begin{thebibliography}{55}%
\makeatletter
\providecommand \@ifxundefined [1]{%
 \@ifx{#1\undefined}
}%
\providecommand \@ifnum [1]{%
 \ifnum #1\expandafter \@firstoftwo
 \else \expandafter \@secondoftwo
 \fi
}%
\providecommand \@ifx [1]{%
 \ifx #1\expandafter \@firstoftwo
 \else \expandafter \@secondoftwo
 \fi
}%
\providecommand \natexlab [1]{#1}%
\providecommand \enquote  [1]{``#1''}%
\providecommand \bibnamefont  [1]{#1}%
\providecommand \bibfnamefont [1]{#1}%
\providecommand \citenamefont [1]{#1}%
\providecommand \href@noop [0]{\@secondoftwo}%
\providecommand \href [0]{\begingroup \@sanitize@url \@href}%
\providecommand \@href[1]{\@@startlink{#1}\@@href}%
\providecommand \@@href[1]{\endgroup#1\@@endlink}%
\providecommand \@sanitize@url [0]{\catcode `\\12\catcode `\$12\catcode
  `\&12\catcode `\#12\catcode `\^12\catcode `\_12\catcode `\%12\relax}%
\providecommand \@@startlink[1]{}%
\providecommand \@@endlink[0]{}%
\providecommand \url  [0]{\begingroup\@sanitize@url \@url }%
\providecommand \@url [1]{\endgroup\@href {#1}{\urlprefix }}%
\providecommand \urlprefix  [0]{URL }%
\providecommand \Eprint [0]{\href }%
\providecommand \doibase [0]{https://doi.org/}%
\providecommand \selectlanguage [0]{\@gobble}%
\providecommand \bibinfo  [0]{\@secondoftwo}%
\providecommand \bibfield  [0]{\@secondoftwo}%
\providecommand \translation [1]{[#1]}%
\providecommand \BibitemOpen [0]{}%
\providecommand \bibitemStop [0]{}%
\providecommand \bibitemNoStop [0]{.\EOS\space}%
\providecommand \EOS [0]{\spacefactor3000\relax}%
\providecommand \BibitemShut  [1]{\csname bibitem#1\endcsname}%
\let\auto@bib@innerbib\@empty
\bibitem [{\citenamefont {Anderson}(1958)}]{Anderson:1958}%
  \BibitemOpen
  \bibfield  {author} {\bibinfo {author} {\bibfnamefont {P.~W.}\ \bibnamefont
  {Anderson}},\ }\href {https://doi.org/10.1103/PhysRev.109.1492} {\bibfield
  {journal} {\bibinfo  {journal} {Phys. Rev.}\ }\textbf {\bibinfo {volume}
  {109}},\ \bibinfo {pages} {1492} (\bibinfo {year} {1958})}\BibitemShut
  {NoStop}%
\bibitem [{\citenamefont {Kramer}\ and\ \citenamefont
  {MacKinnon}(1993)}]{Kramer:1993}%
  \BibitemOpen
  \bibfield  {author} {\bibinfo {author} {\bibfnamefont {B.}~\bibnamefont
  {Kramer}}\ and\ \bibinfo {author} {\bibfnamefont {A.}~\bibnamefont
  {MacKinnon}},\ }\href {https://doi.org/10.1088/0034-4885/56/12/001}
  {\bibfield  {journal} {\bibinfo  {journal} {Rep. Prog. Phys.}\ }\textbf
  {\bibinfo {volume} {56}},\ \bibinfo {pages} {1469} (\bibinfo {year}
  {1993})}\BibitemShut {NoStop}%
\bibitem [{\citenamefont {Evers}\ and\ \citenamefont
  {Mirlin}(2008)}]{Evers:2008}%
  \BibitemOpen
  \bibfield  {author} {\bibinfo {author} {\bibfnamefont {F.}~\bibnamefont
  {Evers}}\ and\ \bibinfo {author} {\bibfnamefont {A.~D.}\ \bibnamefont
  {Mirlin}},\ }\href {https://doi.org/10.1103/RevModPhys.80.1355} {\bibfield
  {journal} {\bibinfo  {journal} {Rev. Mod. Phys.}\ }\textbf {\bibinfo {volume}
  {80}},\ \bibinfo {pages} {1355} (\bibinfo {year} {2008})}\BibitemShut
  {NoStop}%
\bibitem [{\citenamefont {Luck}(1992)}]{Luck:1992}%
  \BibitemOpen
  \bibfield  {author} {\bibinfo {author} {\bibfnamefont {J.-M.}\ \bibnamefont
  {Luck}},\ }\href@noop {} {\emph {\bibinfo {title} {Syst{\`e}mes
  d{\'e}sordonn{\'e}s unidimentionnels}}},\ Collection Al{\'e}a-Saclay\
  (\bibinfo  {publisher} {Commissariat {\`a} l'{\'e}nergie atomique},\ \bibinfo
  {year} {1992})\BibitemShut {NoStop}%
\bibitem [{\citenamefont {Texier}(1999)}]{Texier:1999}%
  \BibitemOpen
  \bibfield  {author} {\bibinfo {author} {\bibfnamefont {C.}~\bibnamefont
  {Texier}},\ }\emph {\bibinfo {title} {{Quelques aspects du transport
  quantique dans les syst{\`e}mes d{\'e}sordonn{\'e}s de basse dimension}}},\
  \href@noop {} {Ph.D. thesis},\ \bibinfo  {school} {Universit{\'e} Paris-Sud}
  (\bibinfo {year} {1999})\BibitemShut {NoStop}%
\bibitem [{\citenamefont {Texier}(2010)}]{Texier:2010}%
  \BibitemOpen
  \bibfield  {author} {\bibinfo {author} {\bibfnamefont {C.}~\bibnamefont
  {Texier}},\ }\emph {\bibinfo {title} {{Disorder, localisation and
  interaction. Quantum transport in metallic networks}}},\ \href
  {https://tel.archives-ouvertes.fr/tel-01091550} {\bibinfo {type}
  {Habilitation {\`a} diriger des recherches}},\ \bibinfo  {school}
  {{Universit{\'e} Paris-Sud}} (\bibinfo {year} {2010})\BibitemShut {NoStop}%
\bibitem [{\citenamefont {Frisch}\ and\ \citenamefont
  {Lloyd}(1960)}]{Frisch:1960}%
  \BibitemOpen
  \bibfield  {author} {\bibinfo {author} {\bibfnamefont {H.~L.}\ \bibnamefont
  {Frisch}}\ and\ \bibinfo {author} {\bibfnamefont {S.~P.}\ \bibnamefont
  {Lloyd}},\ }\href {https://doi.org/10.1103/PhysRev.120.1175} {\bibfield
  {journal} {\bibinfo  {journal} {Phys. Rev.}\ }\textbf {\bibinfo {volume}
  {120}},\ \bibinfo {pages} {1175} (\bibinfo {year} {1960})}\BibitemShut
  {NoStop}%
\bibitem [{\citenamefont {Halperin}(1965)}]{Halperin:1965}%
  \BibitemOpen
  \bibfield  {author} {\bibinfo {author} {\bibfnamefont {B.~I.}\ \bibnamefont
  {Halperin}},\ }\href {https://doi.org/10.1103/PhysRev.139.A104} {\bibfield
  {journal} {\bibinfo  {journal} {Phys. Rev.}\ }\textbf {\bibinfo {volume}
  {139}},\ \bibinfo {pages} {104} (\bibinfo {year} {1965})}\BibitemShut
  {NoStop}%
\bibitem [{\citenamefont {Herbert}\ and\ \citenamefont
  {Jones}(1971)}]{Herbert:1971}%
  \BibitemOpen
  \bibfield  {author} {\bibinfo {author} {\bibfnamefont {D.~C.}\ \bibnamefont
  {Herbert}}\ and\ \bibinfo {author} {\bibfnamefont {R.}~\bibnamefont
  {Jones}},\ }\href {https://doi.org/10.1088/0022-3719/4/10/023} {\bibfield
  {journal} {\bibinfo  {journal} {J. Phys. C}\ }\textbf {\bibinfo {volume}
  {4}},\ \bibinfo {pages} {1145} (\bibinfo {year} {1971})}\BibitemShut
  {NoStop}%
\bibitem [{\citenamefont {Thouless}(1972)}]{Thouless:1972}%
  \BibitemOpen
  \bibfield  {author} {\bibinfo {author} {\bibfnamefont {D.~J.}\ \bibnamefont
  {Thouless}},\ }\href {https://doi.org/10.1088/0022-3719/5/1/010} {\bibfield
  {journal} {\bibinfo  {journal} {J. Phys. C}\ }\textbf {\bibinfo {volume}
  {5}},\ \bibinfo {pages} {77} (\bibinfo {year} {1972})}\BibitemShut {NoStop}%
\bibitem [{\citenamefont {Antsygina}\ \emph {et~al.}(1981)\citenamefont
  {Antsygina}, \citenamefont {Pastur},\ and\ \citenamefont
  {Slyusarev}}]{Antsygina:1981}%
  \BibitemOpen
  \bibfield  {author} {\bibinfo {author} {\bibfnamefont {T.~N.}\ \bibnamefont
  {Antsygina}}, \bibinfo {author} {\bibfnamefont {L.~A.}\ \bibnamefont
  {Pastur}},\ and\ \bibinfo {author} {\bibfnamefont {V.~A.}\ \bibnamefont
  {Slyusarev}},\ }\href {https://doi.org/10.1063/10.0030197} {\bibfield
  {journal} {\bibinfo  {journal} {Sov. J. Low Temp. Phys.}\ }\textbf {\bibinfo
  {volume} {7}},\ \bibinfo {pages} {1} (\bibinfo {year} {1981})}\BibitemShut
  {NoStop}%
\bibitem [{\citenamefont {Lifshits}\ \emph {et~al.}(1988)\citenamefont
  {Lifshits}, \citenamefont {Gredeskul},\ and\ \citenamefont
  {Pastur}}]{Lifshits:1988}%
  \BibitemOpen
  \bibfield  {author} {\bibinfo {author} {\bibfnamefont {I.~M.}\ \bibnamefont
  {Lifshits}}, \bibinfo {author} {\bibfnamefont {S.~A.}\ \bibnamefont
  {Gredeskul}},\ and\ \bibinfo {author} {\bibfnamefont {L.~A.}\ \bibnamefont
  {Pastur}},\ }\href@noop {} {\emph {\bibinfo {title} {Introduction to the
  Theory of Disordered Systems}}}\ (\bibinfo  {publisher} {Wiley, New York},\
  \bibinfo {year} {1988})\BibitemShut {NoStop}%
\bibitem [{\citenamefont {Ovchinnikov}\ and\ \citenamefont
  {Écrikhman}(1977)}]{Ovchinnikov:1977}%
  \BibitemOpen
  \bibfield  {author} {\bibinfo {author} {\bibfnamefont {A.~A.}\ \bibnamefont
  {Ovchinnikov}}\ and\ \bibinfo {author} {\bibfnamefont {N.~S.}\ \bibnamefont
  {Écrikhman}},\ }\href@noop {} {\bibfield  {journal} {\bibinfo  {journal}
  {Zh. Eksp. Teor. Fiz}\ }\textbf {\bibinfo {volume} {73}},\ \bibinfo {pages}
  {650} (\bibinfo {year} {1977})}\BibitemShut {NoStop}%
\bibitem [{\citenamefont {Nieuwenhuizen}(1983)}]{Nieuwenhuizen:1983}%
  \BibitemOpen
  \bibfield  {author} {\bibinfo {author} {\bibfnamefont {T.}~\bibnamefont
  {Nieuwenhuizen}},\ }\href {https://doi.org/10.1016/0378-4371(83)90063-8}
  {\bibfield  {journal} {\bibinfo  {journal} {Physica A}\ }\textbf {\bibinfo
  {volume} {120}},\ \bibinfo {pages} {468} (\bibinfo {year}
  {1983})}\BibitemShut {NoStop}%
\bibitem [{\citenamefont {Bouchaud}\ \emph {et~al.}(1990)\citenamefont
  {Bouchaud}, \citenamefont {Comtet}, \citenamefont {Georges},\ and\
  \citenamefont {Le~Doussal}}]{Bouchaud:1990}%
  \BibitemOpen
  \bibfield  {author} {\bibinfo {author} {\bibfnamefont {J.}~\bibnamefont
  {Bouchaud}}, \bibinfo {author} {\bibfnamefont {A.}~\bibnamefont {Comtet}},
  \bibinfo {author} {\bibfnamefont {A.}~\bibnamefont {Georges}},\ and\ \bibinfo
  {author} {\bibfnamefont {P.}~\bibnamefont {Le~Doussal}},\ }\href
  {https://doi.org/10.1016/0003-4916(90)90043-N} {\bibfield  {journal}
  {\bibinfo  {journal} {Ann. Phys.}\ }\textbf {\bibinfo {volume} {201}},\
  \bibinfo {pages} {285} (\bibinfo {year} {1990})}\BibitemShut {NoStop}%
\bibitem [{\citenamefont {Barnes}\ and\ \citenamefont
  {Luck}(1990)}]{Barnes:1990}%
  \BibitemOpen
  \bibfield  {author} {\bibinfo {author} {\bibfnamefont {C.}~\bibnamefont
  {Barnes}}\ and\ \bibinfo {author} {\bibfnamefont {J.~M.}\ \bibnamefont
  {Luck}},\ }\href {https://doi.org/10.1088/0305-4470/23/10/014} {\bibfield
  {journal} {\bibinfo  {journal} {J. Phys. A}\ }\textbf {\bibinfo {volume}
  {23}},\ \bibinfo {pages} {1717} (\bibinfo {year} {1990})}\BibitemShut
  {NoStop}%
\bibitem [{\citenamefont {Comtet}\ \emph {et~al.}(1995)\citenamefont {Comtet},
  \citenamefont {Desbois},\ and\ \citenamefont {Monthus}}]{Comtet:1995}%
  \BibitemOpen
  \bibfield  {author} {\bibinfo {author} {\bibfnamefont {A.}~\bibnamefont
  {Comtet}}, \bibinfo {author} {\bibfnamefont {J.}~\bibnamefont {Desbois}},\
  and\ \bibinfo {author} {\bibfnamefont {C.}~\bibnamefont {Monthus}},\ }\href
  {https://doi.org/10.1006/aphy.1995.1037} {\bibfield  {journal} {\bibinfo
  {journal} {Ann. Phys. (N. Y.)}\ }\textbf {\bibinfo {volume} {239}},\ \bibinfo
  {pages} {312} (\bibinfo {year} {1995})}\BibitemShut {NoStop}%
\bibitem [{\citenamefont {Hagendorf}\ and\ \citenamefont
  {Texier}(2008)}]{Hagendorf:2008}%
  \BibitemOpen
  \bibfield  {author} {\bibinfo {author} {\bibfnamefont {C.}~\bibnamefont
  {Hagendorf}}\ and\ \bibinfo {author} {\bibfnamefont {C.}~\bibnamefont
  {Texier}},\ }\href {https://doi.org/10.1088/1751-8113/41/40/405302}
  {\bibfield  {journal} {\bibinfo  {journal} {J. Phys. A}\ }\textbf {\bibinfo
  {volume} {41}},\ \bibinfo {pages} {405302} (\bibinfo {year}
  {2008})}\BibitemShut {NoStop}%
\bibitem [{\citenamefont {Comtet}\ \emph {et~al.}(2010)\citenamefont {Comtet},
  \citenamefont {Texier},\ and\ \citenamefont {Tourigny}}]{Comtet:2010}%
  \BibitemOpen
  \bibfield  {author} {\bibinfo {author} {\bibfnamefont {A.}~\bibnamefont
  {Comtet}}, \bibinfo {author} {\bibfnamefont {C.}~\bibnamefont {Texier}},\
  and\ \bibinfo {author} {\bibfnamefont {Y.}~\bibnamefont {Tourigny}},\ }\href
  {https://doi.org/10.1007/s10955-010-0005-x} {\bibfield  {journal} {\bibinfo
  {journal} {J. Stat. Phys.}\ }\textbf {\bibinfo {volume} {140}},\ \bibinfo
  {pages} {426} (\bibinfo {year} {2010})}\BibitemShut {NoStop}%
\bibitem [{\citenamefont {Comtet}\ \emph {et~al.}(2011)\citenamefont {Comtet},
  \citenamefont {Texier},\ and\ \citenamefont {Tourigny}}]{Comtet:2011}%
  \BibitemOpen
  \bibfield  {author} {\bibinfo {author} {\bibfnamefont {A.}~\bibnamefont
  {Comtet}}, \bibinfo {author} {\bibfnamefont {C.}~\bibnamefont {Texier}},\
  and\ \bibinfo {author} {\bibfnamefont {Y.}~\bibnamefont {Tourigny}},\ }\href
  {https://doi.org/10.1007/s10955-011-0351-3} {\bibfield  {journal} {\bibinfo
  {journal} {J.~Stat. Phys.}\ }\textbf {\bibinfo {volume} {145}},\ \bibinfo
  {pages} {1291} (\bibinfo {year} {2011})}\BibitemShut {NoStop}%
\bibitem [{\citenamefont {Comtet}\ \emph {et~al.}(2013)\citenamefont {Comtet},
  \citenamefont {Luck}, \citenamefont {Texier},\ and\ \citenamefont
  {Tourigny}}]{Comtet:2013}%
  \BibitemOpen
  \bibfield  {author} {\bibinfo {author} {\bibfnamefont {A.}~\bibnamefont
  {Comtet}}, \bibinfo {author} {\bibfnamefont {J.-M.}\ \bibnamefont {Luck}},
  \bibinfo {author} {\bibfnamefont {C.}~\bibnamefont {Texier}},\ and\ \bibinfo
  {author} {\bibfnamefont {Y.}~\bibnamefont {Tourigny}},\ }\href
  {https://doi.org/10.1007/s10955-012-0674-8} {\bibfield  {journal} {\bibinfo
  {journal} {J. Stat. Phys.}\ }\textbf {\bibinfo {volume} {150}},\ \bibinfo
  {pages} {13} (\bibinfo {year} {2013})}\BibitemShut {NoStop}%
\bibitem [{\citenamefont {Grabsch}\ \emph {et~al.}(2014)\citenamefont
  {Grabsch}, \citenamefont {Texier},\ and\ \citenamefont
  {Tourigny}}]{Grabsch:2014}%
  \BibitemOpen
  \bibfield  {author} {\bibinfo {author} {\bibfnamefont {A.}~\bibnamefont
  {Grabsch}}, \bibinfo {author} {\bibfnamefont {C.}~\bibnamefont {Texier}},\
  and\ \bibinfo {author} {\bibfnamefont {Y.}~\bibnamefont {Tourigny}},\ }\href
  {https://doi.org/10.1007/s10955-014-0957-3} {\bibfield  {journal} {\bibinfo
  {journal} {J.~Stat. Phys.}\ }\textbf {\bibinfo {volume} {155}},\ \bibinfo
  {pages} {237} (\bibinfo {year} {2014})},\ \Eprint
  {https://arxiv.org/abs/1310.6519v1} {1310.6519v1} \BibitemShut {NoStop}%
\bibitem [{\citenamefont {Dumaz}\ and\ \citenamefont
  {Labb{\'e}}(2020)}]{Dumaz:2020}%
  \BibitemOpen
  \bibfield  {author} {\bibinfo {author} {\bibfnamefont {L.}~\bibnamefont
  {Dumaz}}\ and\ \bibinfo {author} {\bibfnamefont {C.}~\bibnamefont
  {Labb{\'e}}},\ }\href {https://doi.org/10.1007/s00440-019-00920-6} {\bibfield
   {journal} {\bibinfo  {journal} {Probab. Theory Relat. Fields}\ }\textbf
  {\bibinfo {volume} {176}},\ \bibinfo {pages} {353} (\bibinfo {year}
  {2020})}\BibitemShut {NoStop}%
\bibitem [{\citenamefont {Dumaz}\ and\ \citenamefont
  {Labb{\'e}}(2023)}]{Dumaz:2023}%
  \BibitemOpen
  \bibfield  {author} {\bibinfo {author} {\bibfnamefont {L.}~\bibnamefont
  {Dumaz}}\ and\ \bibinfo {author} {\bibfnamefont {C.}~\bibnamefont
  {Labb{\'e}}},\ }\href {https://doi.org/10.1214/22-AOP1591} {\bibfield
  {journal} {\bibinfo  {journal} {Ann. Proba.}\ }\textbf {\bibinfo {volume}
  {51}},\ \bibinfo {pages} {805 } (\bibinfo {year} {2023})}\BibitemShut
  {NoStop}%
\bibitem [{\citenamefont {Dumaz}\ and\ \citenamefont
  {Labbé}(2024)}]{Dumaz:2024b}%
  \BibitemOpen
  \bibfield  {author} {\bibinfo {author} {\bibfnamefont {L.}~\bibnamefont
  {Dumaz}}\ and\ \bibinfo {author} {\bibfnamefont {C.}~\bibnamefont {Labbé}},\
  }\href {https://doi.org/https://doi.org/10.1016/j.jfa.2023.110191} {\bibfield
   {journal} {\bibinfo  {journal} {J. Funct. Anal.}\ }\textbf {\bibinfo
  {volume} {286}},\ \bibinfo {pages} {110191} (\bibinfo {year}
  {2024})}\BibitemShut {NoStop}%
\bibitem [{\citenamefont {Dumaz}\ and\ \citenamefont
  {Labb{\'e}}(2024)}]{Dumaz:2024}%
  \BibitemOpen
  \bibfield  {author} {\bibinfo {author} {\bibfnamefont {L.}~\bibnamefont
  {Dumaz}}\ and\ \bibinfo {author} {\bibfnamefont {C.}~\bibnamefont
  {Labb{\'e}}},\ }\href {https://doi.org/10.1007/s00222-023-01225-1} {\bibfield
   {journal} {\bibinfo  {journal} {Invent. Math.}\ }\textbf {\bibinfo {volume}
  {235}},\ \bibinfo {pages} {345} (\bibinfo {year} {2024})}\BibitemShut
  {NoStop}%
\bibitem [{\citenamefont {Beenakker}(1997)}]{Beenakker:1997}%
  \BibitemOpen
  \bibfield  {author} {\bibinfo {author} {\bibfnamefont {C.~W.~J.}\
  \bibnamefont {Beenakker}},\ }\href
  {https://doi.org/10.1103/RevModPhys.69.731} {\bibfield  {journal} {\bibinfo
  {journal} {Rev. Mod. Phys.}\ }\textbf {\bibinfo {volume} {69}},\ \bibinfo
  {pages} {731} (\bibinfo {year} {1997})},\ \bibinfo {note}
  {\href{https://arxiv.org/abs/cond-mat/9612179}{arXiv:cond-mat/9612179}}\BibitemShut
  {NoStop}%
\bibitem [{\citenamefont {Dorokhov}(1982{\natexlab{a}})}]{Dorokhov:1982}%
  \BibitemOpen
  \bibfield  {author} {\bibinfo {author} {\bibfnamefont {O.}~\bibnamefont
  {Dorokhov}},\ }\href@noop {} {\bibfield  {journal} {\bibinfo  {journal} {JETP
  Lett.}\ }\textbf {\bibinfo {volume} {36}},\ \bibinfo {pages} {318} (\bibinfo
  {year} {1982}{\natexlab{a}})}\BibitemShut {NoStop}%
\bibitem [{\citenamefont {Dorokhov}(1982{\natexlab{b}})}]{Dorokhov:1982b}%
  \BibitemOpen
  \bibfield  {author} {\bibinfo {author} {\bibfnamefont {O.}~\bibnamefont
  {Dorokhov}},\ }\href {https://doi.org/10.1016/0038-1098(82)90304-0}
  {\bibfield  {journal} {\bibinfo  {journal} {Solid State Comm.}\ }\textbf
  {\bibinfo {volume} {44}},\ \bibinfo {pages} {915} (\bibinfo {year}
  {1982}{\natexlab{b}})}\BibitemShut {NoStop}%
\bibitem [{\citenamefont {Dorokhov}(1983)}]{Dorokhov:1983}%
  \BibitemOpen
  \bibfield  {author} {\bibinfo {author} {\bibfnamefont {O.}~\bibnamefont
  {Dorokhov}},\ }\href@noop {} {\bibfield  {journal} {\bibinfo  {journal} {Sov.
  Phys. JETP}\ }\textbf {\bibinfo {volume} {58}},\ \bibinfo {pages} {606}
  (\bibinfo {year} {1983})}\BibitemShut {NoStop}%
\bibitem [{\citenamefont {Dorokhov}(1988)}]{Dorokhov:1988}%
  \BibitemOpen
  \bibfield  {author} {\bibinfo {author} {\bibfnamefont {O.~N.}\ \bibnamefont
  {Dorokhov}},\ }\href {https://doi.org/10.1103/PhysRevB.37.10526} {\bibfield
  {journal} {\bibinfo  {journal} {Phys. Rev. B}\ }\textbf {\bibinfo {volume}
  {37}},\ \bibinfo {pages} {10526} (\bibinfo {year} {1988})}\BibitemShut
  {NoStop}%
\bibitem [{\citenamefont {Mello}\ \emph {et~al.}(1988)\citenamefont {Mello},
  \citenamefont {Pereyra},\ and\ \citenamefont {Kumar}}]{Mello:1988}%
  \BibitemOpen
  \bibfield  {author} {\bibinfo {author} {\bibfnamefont {P.}~\bibnamefont
  {Mello}}, \bibinfo {author} {\bibfnamefont {P.}~\bibnamefont {Pereyra}},\
  and\ \bibinfo {author} {\bibfnamefont {N.}~\bibnamefont {Kumar}},\ }\href
  {https://doi.org/10.1016/0003-4916(88)90169-8} {\bibfield  {journal}
  {\bibinfo  {journal} {Ann. Phys.}\ }\textbf {\bibinfo {volume} {181}},\
  \bibinfo {pages} {290 } (\bibinfo {year} {1988})}\BibitemShut {NoStop}%
\bibitem [{\citenamefont {Mello}\ and\ \citenamefont
  {Kumar}(2004)}]{Mello:2004}%
  \BibitemOpen
  \bibfield  {author} {\bibinfo {author} {\bibfnamefont {P.~A.}\ \bibnamefont
  {Mello}}\ and\ \bibinfo {author} {\bibfnamefont {N.}~\bibnamefont {Kumar}},\
  }\href@noop {} {\emph {\bibinfo {title} {Quantum transport in mesoscopic
  systems -- Complexity and statistical fluctuations}}}\ (\bibinfo  {publisher}
  {Oxford University Press},\ \bibinfo {year} {2004})\BibitemShut {NoStop}%
\bibitem [{\citenamefont {Ossipov}(2018)}]{Ossipov:2018}%
  \BibitemOpen
  \bibfield  {author} {\bibinfo {author} {\bibfnamefont {A.}~\bibnamefont
  {Ossipov}},\ }\href {https://doi.org/10.1103/PhysRevLett.121.076601}
  {\bibfield  {journal} {\bibinfo  {journal} {Phys. Rev. Lett.}\ }\textbf
  {\bibinfo {volume} {121}},\ \bibinfo {pages} {076601} (\bibinfo {year}
  {2018})}\BibitemShut {NoStop}%
\bibitem [{\citenamefont {Brouwer}\ \emph {et~al.}(2000)\citenamefont
  {Brouwer}, \citenamefont {Mudry},\ and\ \citenamefont
  {Furusaki}}]{Brouwer:2000}%
  \BibitemOpen
  \bibfield  {author} {\bibinfo {author} {\bibfnamefont {P.~W.}\ \bibnamefont
  {Brouwer}}, \bibinfo {author} {\bibfnamefont {C.}~\bibnamefont {Mudry}},\
  and\ \bibinfo {author} {\bibfnamefont {A.}~\bibnamefont {Furusaki}},\ }\href
  {https://doi.org/10.1103/PhysRevLett.84.2913} {\bibfield  {journal} {\bibinfo
   {journal} {Phys. Rev. Lett.}\ }\textbf {\bibinfo {volume} {84}},\ \bibinfo
  {pages} {2913} (\bibinfo {year} {2000})}\BibitemShut {NoStop}%
\bibitem [{\citenamefont {Titov}\ \emph {et~al.}(2001)\citenamefont {Titov},
  \citenamefont {Brouwer}, \citenamefont {Furusaki},\ and\ \citenamefont
  {Mudry}}]{Titov:2001}%
  \BibitemOpen
  \bibfield  {author} {\bibinfo {author} {\bibfnamefont {M.}~\bibnamefont
  {Titov}}, \bibinfo {author} {\bibfnamefont {P.~W.}\ \bibnamefont {Brouwer}},
  \bibinfo {author} {\bibfnamefont {A.}~\bibnamefont {Furusaki}},\ and\
  \bibinfo {author} {\bibfnamefont {C.}~\bibnamefont {Mudry}},\ }\href
  {https://doi.org/10.1103/PhysRevB.63.235318} {\bibfield  {journal} {\bibinfo
  {journal} {Phys. Rev. B}\ }\textbf {\bibinfo {volume} {63}},\ \bibinfo
  {pages} {235318} (\bibinfo {year} {2001})}\BibitemShut {NoStop}%
\bibitem [{\citenamefont {Grabsch}\ and\ \citenamefont
  {Texier}(2016)}]{Grabsch:2016b}%
  \BibitemOpen
  \bibfield  {author} {\bibinfo {author} {\bibfnamefont {A.}~\bibnamefont
  {Grabsch}}\ and\ \bibinfo {author} {\bibfnamefont {C.}~\bibnamefont
  {Texier}},\ }\href {https://doi.org/10.1209/0295-5075/116/17004} {\bibfield
  {journal} {\bibinfo  {journal} {Europhys. Lett.}\ }\textbf {\bibinfo {volume}
  {116}},\ \bibinfo {pages} {17004} (\bibinfo {year} {2016})}\BibitemShut
  {NoStop}%
\bibitem [{\citenamefont {Grabsch}(9097)}]{Grabsch:2018}%
  \BibitemOpen
  \bibfield  {author} {\bibinfo {author} {\bibfnamefont {A.}~\bibnamefont
  {Grabsch}},\ }\emph {\bibinfo {title} {Random matrix theory in statistical
  physics: quantum scattering and disordered systems}},\ \href@noop {} {Ph.D.
  thesis},\ \bibinfo  {school} {Universit\'e Paris Saclay} (\bibinfo {year}
  {2018,
  \href{https://tel.archives-ouvertes.fr/tel-01849097}{https://tel.archives-ouvertes.fr/tel-01849097}})\BibitemShut
  {NoStop}%
\bibitem [{\citenamefont {Allez}\ and\ \citenamefont
  {Dumaz}(2015)}]{Allez:2015}%
  \BibitemOpen
  \bibfield  {author} {\bibinfo {author} {\bibfnamefont {R.}~\bibnamefont
  {Allez}}\ and\ \bibinfo {author} {\bibfnamefont {L.}~\bibnamefont {Dumaz}},\
  }\href {https://doi.org/10.1007/s10955-015-1258-1} {\bibfield  {journal}
  {\bibinfo  {journal} {J. Stat. Phys.}\ }\textbf {\bibinfo {volume} {160}},\
  \bibinfo {pages} {681} (\bibinfo {year} {2015})}\BibitemShut {NoStop}%
\bibitem [{\citenamefont {Fyodorov}\ and\ \citenamefont
  {Le~Doussal}(2020)}]{Fyodorov:2020}%
  \BibitemOpen
  \bibfield  {author} {\bibinfo {author} {\bibfnamefont {Y.~V.}\ \bibnamefont
  {Fyodorov}}\ and\ \bibinfo {author} {\bibfnamefont {P.}~\bibnamefont
  {Le~Doussal}},\ }\href {https://doi.org/10.1007/s10955-020-02522-2}
  {\bibfield  {journal} {\bibinfo  {journal} {J. Stat. Phys.}\ }\textbf
  {\bibinfo {volume} {179}},\ \bibinfo {pages} {176} (\bibinfo {year}
  {2020})}\BibitemShut {NoStop}%
\bibitem [{\citenamefont {Fyodorov}\ \emph {et~al.}(2026)\citenamefont
  {Fyodorov}, \citenamefont {Le~Doussal},\ and\ \citenamefont
  {Ossipov}}]{Fyodorov:2026a}%
  \BibitemOpen
  \bibfield  {author} {\bibinfo {author} {\bibfnamefont {Y.~V.}\ \bibnamefont
  {Fyodorov}}, \bibinfo {author} {\bibfnamefont {P.}~\bibnamefont
  {Le~Doussal}},\ and\ \bibinfo {author} {\bibfnamefont {A.}~\bibnamefont
  {Ossipov}},\ }\href {https://doi.org/10.1063/5.0351290} {\bibfield  {journal}
  {\bibinfo  {journal} {J. Math. Phys.}\ }\textbf {\bibinfo {volume} {67}},\
  \bibinfo {pages} {103301} (\bibinfo {year} {2026})}\BibitemShut {NoStop}%
\bibitem [{\citenamefont {Fyodorov}\ and\ \citenamefont
  {Doussal}(2026)}]{Fyodorov:2026}%
  \BibitemOpen
  \bibfield  {author} {\bibinfo {author} {\bibfnamefont {Y.~V.}\ \bibnamefont
  {Fyodorov}}\ and\ \bibinfo {author} {\bibfnamefont {P.~L.}\ \bibnamefont
  {Doussal}},\ }\bibfield  {journal} {\bibinfo  {journal} {arXiv preprint
  arXiv:2610.27154}\ }\href {https://doi.org/10.48550/arXiv.2610.27154}
  {10.48550/arXiv.2610.27154} (\bibinfo {year} {2026})\BibitemShut {NoStop}%
\bibitem [{\citenamefont {Fyodorov}(2026)}]{Fyodorov:2026b}%
  \BibitemOpen
  \bibfield  {author} {\bibinfo {author} {\bibfnamefont {Y.~V.}\ \bibnamefont
  {Fyodorov}},\ }\bibfield  {journal} {\bibinfo  {journal} {arXiv preprint
  arXiv:2611.21067}\ }\href {https://doi.org/10.48550/arXiv.2611.21067}
  {10.48550/arXiv.2611.21067} (\bibinfo {year} {2026})\BibitemShut {NoStop}%
\bibitem [{\citenamefont {Lacroix}(1986)}]{Lacroix:1986}%
  \BibitemOpen
  \bibfield  {author} {\bibinfo {author} {\bibfnamefont {J.}~\bibnamefont
  {Lacroix}},\ }in\ \href {https://doi.org/10.1007/BFb0076846} {\emph {\bibinfo
  {booktitle} {Lyapunov Exponents}}},\ \bibinfo {editor} {edited by\ \bibinfo
  {editor} {\bibfnamefont {L.}~\bibnamefont {Arnold}}\ and\ \bibinfo {editor}
  {\bibfnamefont {V.}~\bibnamefont {Wihstutz}}}\ (\bibinfo  {publisher}
  {Springer Berlin Heidelberg},\ \bibinfo {address} {Berlin, Heidelberg},\
  \bibinfo {year} {1986})\ pp.\ \bibinfo {pages} {258--264}\BibitemShut
  {NoStop}%
\bibitem [{\citenamefont {Molinari}(2013)}]{Molinari:2013}%
  \BibitemOpen
  \bibfield  {author} {\bibinfo {author} {\bibfnamefont {L.~G.}\ \bibnamefont
  {Molinari}},\ }\href {https://doi.org/10.1088/1751-8113/46/25/254004}
  {\bibfield  {journal} {\bibinfo  {journal} {J. Phys. A: Math. Theor.}\
  }\textbf {\bibinfo {volume} {46}},\ \bibinfo {pages} {254004} (\bibinfo
  {year} {2013})}\BibitemShut {NoStop}%
\bibitem [{\citenamefont {Derrida}\ \emph {et~al.}(1987)\citenamefont
  {Derrida}, \citenamefont {Mecheri},\ and\ \citenamefont
  {Pichard}}]{Derrida:1987}%
  \BibitemOpen
  \bibfield  {author} {\bibinfo {author} {\bibfnamefont {B.}~\bibnamefont
  {Derrida}}, \bibinfo {author} {\bibfnamefont {K.}~\bibnamefont {Mecheri}},\
  and\ \bibinfo {author} {\bibfnamefont {J.}~\bibnamefont {Pichard}},\ }\href
  {https://doi.org/10.1051/jphys:01987004805073300} {\bibfield  {journal}
  {\bibinfo  {journal} {J. Phys. France}\ }\textbf {\bibinfo {volume} {48}},\
  \bibinfo {pages} {733} (\bibinfo {year} {1987})}\BibitemShut {NoStop}%
\bibitem [{\citenamefont {R{\"o}mer}\ and\ \citenamefont
  {Schulz-Baldes}(2010)}]{Romer:2010}%
  \BibitemOpen
  \bibfield  {author} {\bibinfo {author} {\bibfnamefont {R.~A.}\ \bibnamefont
  {R{\"o}mer}}\ and\ \bibinfo {author} {\bibfnamefont {H.}~\bibnamefont
  {Schulz-Baldes}},\ }\href {https://doi.org/10.1007/s10955-010-9986-8}
  {\bibfield  {journal} {\bibinfo  {journal} {J. Stat. Phys.}\ }\textbf
  {\bibinfo {volume} {140}},\ \bibinfo {pages} {122} (\bibinfo {year}
  {2010})}\BibitemShut {NoStop}%
\bibitem [{\citenamefont {Ludwig}\ \emph {et~al.}(2013)\citenamefont {Ludwig},
  \citenamefont {Schulz-Baldes},\ and\ \citenamefont {Stolz}}]{Ludwig:2013}%
  \BibitemOpen
  \bibfield  {author} {\bibinfo {author} {\bibfnamefont {A.~W.~W.}\
  \bibnamefont {Ludwig}}, \bibinfo {author} {\bibfnamefont {H.}~\bibnamefont
  {Schulz-Baldes}},\ and\ \bibinfo {author} {\bibfnamefont {M.}~\bibnamefont
  {Stolz}},\ }\href {https://doi.org/10.1007/s10955-013-0764-2} {\bibfield
  {journal} {\bibinfo  {journal} {J. Stat. Phys.}\ }\textbf {\bibinfo {volume}
  {152}},\ \bibinfo {pages} {275} (\bibinfo {year} {2013})}\BibitemShut
  {NoStop}%
\bibitem [{\citenamefont {Oseledec}(1968)}]{Oseledec:1968}%
  \BibitemOpen
  \bibfield  {author} {\bibinfo {author} {\bibfnamefont {V.~I.}\ \bibnamefont
  {Oseledec}},\ }\href@noop {} {\bibfield  {journal} {\bibinfo  {journal} {Tr.
  Mosk. Mat. Obs.}\ }\textbf {\bibinfo {volume} {19}},\ \bibinfo {pages} {179}
  (\bibinfo {year} {1968})}\BibitemShut {NoStop}%
\bibitem [{\citenamefont {Kaimanovich}(1989)}]{Kaimanovich:1989}%
  \BibitemOpen
  \bibfield  {author} {\bibinfo {author} {\bibfnamefont {V.~A.}\ \bibnamefont
  {Kaimanovich}},\ }\href {https://doi.org/10.1007/BF01840421} {\bibfield
  {journal} {\bibinfo  {journal} {J. Sov. Math.}\ }\textbf {\bibinfo {volume}
  {47}},\ \bibinfo {pages} {2387} (\bibinfo {year} {1989})}\BibitemShut
  {NoStop}%
\bibitem [{SM()}]{SM}%
  \BibitemOpen
  \href@noop {} {\bibinfo {title} {See the {Supplemental}
  {Material}}}\BibitemShut {NoStop}%
\bibitem [{\citenamefont {Dieci}\ and\ \citenamefont
  {Vleck}(1995)}]{Dieci:1995}%
  \BibitemOpen
  \bibfield  {author} {\bibinfo {author} {\bibfnamefont {L.}~\bibnamefont
  {Dieci}}\ and\ \bibinfo {author} {\bibfnamefont {E.~S.~V.}\ \bibnamefont
  {Vleck}},\ }\href {https://doi.org/10.1016/0168-9274(95)00033-Q} {\bibfield
  {journal} {\bibinfo  {journal} {Appl. Numer. Math.}\ }\textbf {\bibinfo
  {volume} {17}},\ \bibinfo {pages} {275 } (\bibinfo {year} {1995})},\ \bibinfo
  {note} {special Issue on Numerical Methods for Ordinary Differential
  Equations}\BibitemShut {NoStop}%
\bibitem [{\citenamefont {Dieci}\ \emph {et~al.}(2010)\citenamefont {Dieci},
  \citenamefont {Jolly},\ and\ \citenamefont {Vleck}}]{Dieci:2010}%
  \BibitemOpen
  \bibfield  {author} {\bibinfo {author} {\bibfnamefont {L.}~\bibnamefont
  {Dieci}}, \bibinfo {author} {\bibfnamefont {M.~S.}\ \bibnamefont {Jolly}},\
  and\ \bibinfo {author} {\bibfnamefont {E.~S.~V.}\ \bibnamefont {Vleck}},\
  }\href {https://doi.org/10.1115/1.4002088} {\bibfield  {journal} {\bibinfo
  {journal} {J. Comput. Nonlinear Dynam.}\ }\textbf {\bibinfo {volume} {6}},\
  \bibinfo {pages} {011003} (\bibinfo {year} {2010})}\BibitemShut {NoStop}%
\bibitem [{Note1()}]{Note1}%
  \BibitemOpen
  \bibinfo {note} {Since $Q$ is unitary, $Q^\dagger Q = \protect \mathds {1}_N$
  and thus $Q^\dagger \partial _x Q =- (Q^\dagger \partial _x Q)^\dagger $, so
  that $\protect \mathrm {Re}[(Q^\dagger \partial _x Q)_{nn}] =
  0$.}\BibitemShut {Stop}%
\bibitem [{\citenamefont {Bleher}\ and\ \citenamefont
  {Dea{\~n}o}(2016)}]{Bleher:2016}%
  \BibitemOpen
  \bibfield  {author} {\bibinfo {author} {\bibfnamefont {P.}~\bibnamefont
  {Bleher}}\ and\ \bibinfo {author} {\bibfnamefont {A.}~\bibnamefont
  {Dea{\~n}o}},\ }\href {https://doi.org/10.1142/S2010326316500040} {\bibfield
  {journal} {\bibinfo  {journal} {Random Matrices: Theory Appl.}\ }\textbf
  {\bibinfo {volume} {05}},\ \bibinfo {pages} {1650004} (\bibinfo {year}
  {2016})}\BibitemShut {NoStop}%
\end{thebibliography}

%

\appendix
\setcounter{equation}{0}
\setcounter{figure}{0}
\setcounter{table}{0}
\renewcommand{\theequation}{A\arabic{equation}}


\vspace{12pt}
\noindent\hrulefill \hspace{24pt} {\bf End Matter} \hspace{24pt} \hrulefill
\vspace{12pt}

\emph{Gradient and Laplace operators with respect to matrices.---} To write the Fokker-Planck operator in the compact form~(\ref{eq:FokkerPlanck}), we introduce the differential operators acting on functions of matrix argument. First, we define the matrices $\dmat{ij}$ whose elements are all zero, except at position $(i,j)$ which contains a $1$: $(\dmat{ij})_{k,l} = \delta_{i,k} \delta_{j,l}$. We define the gradient with respect to a Hermitian matrix $H$ by separating the independent real variables,
\begin{multline}
  \nabla_H = \sum_{k} \dmat{kk} \frac{\partial}{\partial {M_{kk}}}
  + \frac{1}{2} \sum_{k<l} \Bigg[
  (\dmat{kl} + \dmat{lk}) \frac{\partial}{\partial {\mathrm{Re} M_{kl}}}
  \\
    + \I  (\dmat{kl} - \dmat{lk}) \frac{\partial}{\partial {\mathrm{Im} M_{lk}}}
    \Bigg]
    \:.
\end{multline}
Similarly, we define the divergence operator acting on a matrix $F(M)$ as
\begin{multline}
  \Div_H [F(H)]
  = \sum_{k} \dep{F_{kk}}{H_{kk}}
  \\
  + \sum_{k<l} \Bigg[
  \dep{\mathrm{Re} F_{kl}}{\mathrm{Re} H_{kl}}
  +  \dep{\mathrm{Im} F_{kl}}{\mathrm{Im} H_{kl}}
  \Bigg]
  \:.
\end{multline}
From these definitions, we deduce the Laplace operator as
\begin{multline}
  \Delta_H = \Div_H [\nabla_H]
  = \sum_{k} \depn{}{M_{kk}}{2}
  \\
  + \frac{1}{2} \sum_{k<l} \Bigg[
  \depn{}{\mathrm{Re} H_{kl}}{2}
  +  \depn{}{\mathrm{Im} H_{kl}}{2}
  \Bigg]
  \:.
\end{multline}
Similarly, for a complex non-Hermitian matrix $M$, we define
\begin{equation}
  \nabla_M = \sum_{k,l} \dmat{kl} \left(
    \dep{}{\mathrm{Re} M_{kl}}
    + \I  \dep{}{\mathrm{Re} M_{kl}}
  \right)
  \:,
\end{equation}
\begin{equation}
  \Div_M[F(M)]
  = \sum_{k,l} \Bigg[
  \dep{\mathrm{Re} F_{kl}}{\mathrm{Re} M_{kl}}
  +  \dep{\mathrm{Im} F_{kl}}{\mathrm{Im} M_{kl}}
  \Bigg]
  \:,
\end{equation}
\begin{equation}
  \Delta_M = \Div_M[ \nabla_M]
  = \sum_{k,l} \Bigg[
  \depn{}{\mathrm{Re} M_{kl}}{2}
  +  \depn{}{\mathrm{Im} M_{kl}}{2}
  \Bigg]
  \:.
\end{equation}
These are the definitions used to write the Fokker-Planck operator~(\ref{eq:FokkerPlanck}).

\emph{Solution of the stationary Fokker-Planck equation.---}
Consider the function
\begin{equation}
  f(X_1,X_2,H) = \e^{- \frac{1}{T} \Tr[H X_1^2 - X_1 H^2 + \frac{H^3}{3} + E H + H X_2 X_2^\dagger]}
  \:,
\end{equation}
where $X_1$ and $H$ are $k \times k$ Hermitian matrices and $X_2$ a $k \times (N-k)$ complex matrix. Writing these matrices in terms of their independent real components, we compute
\begin{multline}
  T \Delta_{X_1} f + \frac{T}{2} \Delta_{X_2} f
  = f \Bigg\lbrace
  \frac{1}{T} \Tr[(X_1 H + H X_1 - H^2)^2]
  \\
  + \frac{2}{T} \Tr[H X_2 X_2^\dagger H]
  - 2 N \Tr[H]
  \Bigg\rbrace
  \:.
\end{multline}
Similarly,
\begin{multline}
  \Div_{X_1}[(X_1^2 + E - Y_2 Y_2^\dagger) f]
  = 2k \Tr[X_1] f
  \\
  - \frac{f}{T} \Tr[(X_1^2 + E - X_2 X_2^\dagger)(X_1 H + H X_1 - H^2)]
  \:,
\end{multline}
\begin{multline}
  \Div_{X_2}[X_1 X_2 f]
  = 2(N-k) \Tr[X_1] f
  \\
  - \frac{f}{T} \Tr[(X_1 H + H X_1) X_2 X_2^\dagger]
  \:.
\end{multline}
Furthermore, noticing that
\begin{equation}
  \nabla_H f = - \frac{f}{T} (X_1^2 - X_1 H - H X_1 + H^2 + E + X_2 X_2^\dagger)
  \:,
\end{equation}
we can write the action of the Fokker-Planck
operator~(\ref{eq:FokkerPlanck}) as
\begin{multline}
  \mathcal{L}^\dagger f = \Tr[(X_1 H + H X_1 - H^2) \nabla_H f]
  \\
  + 2(2N-k) f \Tr[X_1] - 2 N f \Tr[H]
  \:.
\end{multline}
This expression can be further simplified by using that
\begin{multline}
  \Tr[(X_1 H + H X_1 - H^2) \nabla_H f]
  =  - 2 k f \Tr[X_1 - H] 
  \\
  + \Div_H [(X_1 H + H X_1 - H^2) f]
  \:,
\end{multline}
so that
\begin{multline}
  \mathcal{L}^\dagger f =
  \Div_H [(X_1 H + H X_1 - H^2) f]
  \\
  + 2(N-k) f \Tr[2X_1 - H]
  \:.
\end{multline}
The last term can be rewritten by using that $\nabla_H (\det H)^\alpha = \alpha (\det H)^{\alpha} H^{-1}$, so that
\begin{multline}
  \Tr[(X_1 H + H X_1 - H^2) \nabla_H (\det H)^{2(N-k)}](\det H)^{-2(N-k)}
  \\
  = 2(N-k) \Tr[2X_1 - H]
  \:.
\end{multline}
Therefore, introducing now the function
$g(X_1,X_2,H) = (\det H)^{2(N-k)} f(X_1,X_2,H)$, we have
\begin{equation}
  \label{eq:FPasDiv}
  \mathcal{L}^\dagger g =
  \Div_H[(X_1 H + H X_1 - H^2) g]
  \:.
\end{equation}

The last step to verify that~(\ref{eq:JointStatioDistr}) solves the stationary Fokker-Planck equation is to check that the integral of the divergence in~(\ref{eq:FPasDiv}) over the positive Hermitian matrices $H$ vanishes. Performing the integral using the divergence theorem, we obtain
\begin{equation}
  \label{eq:IntegDiv}
  \int_{H>0} \mathcal{L}^\dagger g
  = \int_{\partial(H > 0)} g \Tr[(X_1 H + H X_1 - H^2) N(H)] \dd H
  \:,
\end{equation}
where $\partial(H>0)$ denotes the boundary of the domain $H>0$, i.e. it corresponds to Hermitian matrices with one zero eigenvalues and $N-1$ positive nonzero eigenvalues, with $N(H)$ a matrix normal to the boundary $\partial(H>0)$ pointing towards the outside of the region $H>0$. For $k<N$, $g$ vanishes on this domain since $\det H$ vanishes. For $k=N$, we need to express the normal matrix $N(H)$. Let us denote $v$ the eigenvector of $H$ associated to the zero eigenvalue, $H v = v^\dagger H = 0$. Adding the rank one matrix $- \varepsilon v v^\dagger$ changes the zero eigenvalue into $-\varepsilon$ but does not affect the other ones. Hence, for any $\varepsilon > 0$, this makes the matrix $H - \varepsilon v v^\dagger$ leave the domain $H \geq 0$. Therefore, $N(H) \propto - v v^\dagger$ and $\Tr[(X_1 H + H X_1 - H^2) N(H)] \propto v^\dagger(X_1 H + H X_1 - H^2) v = 0 $ since $v$ is an eigenvector of $H$ with eigenvalue $0$. Finally, since the operator
$\mathcal{L}^\dagger$~(\ref{eq:FokkerPlanck}) is linear and does not act on $H$,
\begin{equation}
  \mathcal{L}^\dagger \int_{H>0}  g(X_1, X_2, H) \dd H = 0
  \:.
\end{equation}
This shows that the distribution~(\ref{eq:JointStatioDistr}) solves the stationary Fokker-Planck equation.

\emph{Scaling function $\Phi$ from the Painlevé I equation.---} For
completeness, we recall here the definition of the scaling function $\Phi$ of Eq.~(\ref{eq:LocLengthLargeN}) given in Ref.~\cite{Fyodorov:2026}. It is defined from the so-called \textit{tritronquée} solution $y_1$ of the Painlevé~I equation
\begin{equation}
  y_1''(x) = 6 y_1(x)^2 + x
  \:,
\end{equation}
with the asymptotic behavior~\cite{Bleher:2016}
\begin{equation}
  y_1(x) \underset{|x| \to \infty}{\simeq}  \sqrt{\frac{-x}{6}}
  \:,
  \quad
  - \frac{\pi}{5} < \mathrm{arg}(x) < \frac{7 \pi }{5}
  \:.
\end{equation}
From this solution, the scaling function $\Phi$ is defined as~\cite{Fyodorov:2026}
\begin{equation}
  \Phi'(x) = -\frac{2^{4/5}}{\pi} \mathrm{Im}[y_1(2^{7/5}x)]
  \:,
\end{equation}
and $\Phi(x) = \int_{-\infty}^x \Phi'(u) \dd u$.

\end{document}


\setcounter{equation}{0}
\setcounter{figure}{0}
\setcounter{table}{0}
\setcounter{page}{1}
\makeatletter
\renewcommand{\theequation}{S\arabic{equation}}
\renewcommand{\thefigure}{S\arabic{figure}}
\renewcommand{\bibnumfmt}[1]{[S#1]}
\renewcommand{\citenumfont}[1]{S#1}

\title{Supplemental Material for\texorpdfstring{\\}{}Exact localisation length for the multichannel disordered Schrödinger equation}

\author{Aur\'elien Grabsch}

\maketitle

\tableofcontents

\section{The Fokker-Planck equation}

We start from the joint stochastic equations satisfied by the blocks of the matrix $X$,
\begin{equation}
  \label{eq:JointEqsMatX}
  \partial_x X_1 = - X_1^2 + X_2 X_2^\dagger - E \un_k + W_1
  \:,
  \quad
  \partial_x X_2 = - 2 X_1 X_2 + W_2
  \:,
\end{equation}
where $X_1$ has size $k \times k$, $X_2$ size $k \times (N-k)$, $W_1$ and $W_2$ are the blocks of the Hermitian noise matrix $W$,
\begin{equation}
  \label{eq:matW}
  W =
  \begin{pmatrix}
    W_1 & W_2
    \\
    W_2^\dagger & W_3
  \end{pmatrix}
  \:,
  \quad
  P[W] \propto \exp \left[ - \frac{1}{4T}\int \Tr[W^\dagger(x) W(x)] \dd x \right]
  \:.
\end{equation}
We denote
\begin{equation}
  (W_1)_{ii} = \sqrt{2T} \: \eta_{ii}
  \:,
  \quad
  (W_1)_{ij \neq i} = \sqrt{T} (\eta_{ij} + \I \xi_{ij} )
  \:,
  \quad
  (W_2)_{ij} = \sqrt{T} (\tilde\eta_{ij} + \I \tilde\xi_{ij} )
  \:,
\end{equation}
with $\eta_{ij} = \eta_{ji}$ and $\xi_{ij} = -\xi_{ji}$ real. Inserting these expressions into the distribution of $W$~(\ref{eq:matW}), we obtain by regrouping the independent real elements
\begin{equation}
  P[\{ \eta, \xi, \tilde\eta, \tilde\xi \}]
  \propto
  \exp \left[
    - \frac{1}{2} \int \dd x \left(
      \sum_i \eta_{ii}^2 + \sum_{i<j}( \eta_{ij}^2 + \xi_{ij}^2)
      + \sum_{ij} ( \tilde\eta_{ij}^2 + \tilde\xi_{ij}^2)
    \right)
  \right]
  \:.
\end{equation}
Thus, $\eta$, $\xi$, $\tilde\eta$, $\tilde\xi$ are independent Gaussian white noises, with unit variance.

The joint distribution $p_t(X_1,X_2)$ thus obeys the Fokker-Planck equation $\partial_t p_t = \mathcal{L}^\dagger p_t$, where the operator $\mathcal{L}^\dagger$ can be expressed by considering the independent real elements of $X_1$ and $X_2$ as
\begin{multline}
  \mathcal{L}^\dagger p
  = - \sum_{i} \dep{}{(X_1)_{ii}}[(- X_1^2 +X_2 X_2^\dagger - E)_{ii}p]
  \\
  - \sum_{i<j} \left\lbrace
    \dep{}{\mathrm{Re}(X_1)_{ij}}[\mathrm{Re}(- X_1^2 +X_2 X_2^\dagger )_{ij}p]
    + \dep{}{\mathrm{Im}(X_1)_{ij}}[\mathrm{Im}(- X_1^2 +X_2 X_2^\dagger)_{ij}p]
  \right\rbrace
  \\
  - \sum_{i,j} \left\lbrace
    \dep{}{\mathrm{Re}(X_2)_{ij}}[\mathrm{Re}( - 2 X_1 X_2 )_{ij}p]
    + \dep{}{\mathrm{Im}(X_2)_{ij}}[\mathrm{Im}(-2 X_1 X_2)_{ij}p]
  \right\rbrace
  \\
  + T \sum_i \depn{p}{(X_1)_{ii}}{2}
  + \frac{T}{2} \sum_{i<j}  \left\lbrace
    \depn{p}{\mathrm{Re}(X_1)_{ij}}{2}
    + \depn{p}{\mathrm{Im}(X_1)_{ij}}{2}
  \right\rbrace
  + \frac{T}{2} \sum_{i,j} \left\lbrace
    \depn{p}{\mathrm{Re}(X_2)_{ij}}{2}
    + \depn{p}{\mathrm{Im}(X_2)_{ij}}{2}
  \right\rbrace
  \:.
\end{multline}
Using the definitions of the divergence and Laplace operators for matrix arguments given in the End Matter, we can write this expression in a compact form
\begin{equation}
  \label{eq:FokkerPlanckCompact}
  \mathcal{L}^\dagger p
  = \Div_{X_1}[(X_1^2 - X_2 X_2^\dagger + E \un_k) p]
  + \Div_{X_2}[ 2 X_1 X_2 p]
  + T \Delta_{X_1} p + \frac{T}{2} \Delta_{X_2} p
  \:.
\end{equation}
This is Eq.~(18) of the main text.

\section{The Lyapunov exponents}
\label{sec:Lyaps}

As shown in the main text, the Lyapunov exponents are obtained from the average of $\Tr[X_1]$ with respect to the stationary distribution
\begin{equation}
  \label{eq:JointStatioDistrSM}
  \pst(X_1,X_2) = \frac{1}{\mathcal{N}}
  \int_{H > 0} \dd H \:
  (\det H)^{2(N-k)}
  \e^{- \frac{1}{T} \Tr[H X_1^2 - X_1 H^2 + \frac{H^3}{3} + E H + H X_2 X_2^\dagger]}
  \:.
\end{equation}
Let us first compute the marginal of $X_1$. Making the change of variables $X_2 = H^{-1/2} Y_2$, whose Jacobian is~\cite{Mathai:1997}
\begin{equation}
  \dd X_2 = (\det H)^{-(N-k)} \dd Y_2
  \:,
\end{equation}
since $X_2$ has dimension $k \times (N-k)$. We obtain
\begin{equation}
  p(X_1) = \frac{1}{\mathcal{N}}
  \int_{H>0} \dd H \: (\det H)^{N-k}
  \e^{- \frac{1}{T} \Tr[H X_1^2 - X_1 H^2 + \frac{H^3}{3} + E H]}
  \int \dd Y_2 \: \e^{- \frac{1}{T} \Tr[Y_2 Y_2^\dagger]}
  \:.
\end{equation}
Since the integral over $Y_2$ does not depend on $X_1$ or $H$, we get
\begin{equation}
  p(X_1) = \frac{1}{\tilde{\mathcal{N}}}
  \int_{H>0} \dd H \: (\det H)^{N-k}
  \e^{- \frac{1}{T} \Tr[H X_1^2 - X_1 H^2 + \frac{H^3}{3} + E H]}
  \:,
\end{equation}
with a different normalization constant $\tilde{\mathcal{N}}$. We can thus write
\begin{equation}
  \moyst{\Tr[X_1]}
  = \frac{
    \displaystyle
    \int_{H>0} \dd H \int \dd X_1 \:
    \Tr[X_1] (\det H)^{N-k}
    \e^{- \frac{1}{T} \Tr[H X_1^2 - X_1 H^2 + \frac{H^3}{3} + E H]}
  }{
    \displaystyle
    \int_{H>0} \dd H \int \dd X_1 \:
    (\det H)^{N-k}
    \e^{- \frac{1}{T} \Tr[H X_1^2 - X_1 H^2 + \frac{H^3}{3} + E H]}
  }
  \:.
\end{equation}
The integral over $X_1$ is Gaussian. To simplify it we make the translation $X_1 = \frac{H}{2} + M$, which gives
\begin{equation}
  \moyst{\Tr[X_1]}
  = \frac{
    \displaystyle
    \int_{H>0} \dd H (\det H)^{N-k}
    \e^{- \frac{1}{T} \Tr[\frac{H^3}{12} + E H]}
    \int \dd M \:
    \Tr\left[\frac{H}{2} + M\right]
    \e^{- \frac{1}{T} \Tr[H M^2]}
  }{
    \displaystyle
    \int_{H>0} \dd H (\det H)^{N-k}
    \e^{- \frac{1}{T} \Tr[\frac{H^3}{12} + E H]}
    \int \dd M \:
    \e^{- \frac{1}{T} \Tr[H M^2]}
  }
  \:.
\end{equation}
The integral with the term $\Tr[M]$ vanishes by symmetry, hence
\begin{equation}
  \moyst{\Tr[X_1]}
  = \frac{1}{2}
  \frac{
    \displaystyle
    \int_{H>0} \dd H \: \Tr[H] (\det H)^{N-k}
    \e^{- \frac{1}{T} \Tr[\frac{H^3}{12} + E H]}
    \int \dd M \:
    \e^{- \frac{1}{T} \Tr[H M^2]}
  }{
    \displaystyle
    \int_{H>0}  (\det H)^{N-k}
    \e^{- \frac{1}{T} \Tr[\frac{H^3}{12} + E H]}
    \int \dd M \:
    \e^{- \frac{1}{T} \Tr[H M^2]}
  }
  \:.
\end{equation}
To compute these integrals, we make a change of variables to the eigenvalues and eigenvectors of $H$~\cite{Mathai:1997},
\begin{equation}
  H = V \: \mathrm{Diag}(h_1, \ldots, h_k) \: V^\dagger
  \:,
  \quad
  \dd H = \Delta(h)^2 \dd^k h \: \dd \mu(V)
  \:,
  \quad
  \Delta(h) = \prod_{i<j} (h_j - h_i)
  \:,
\end{equation}
and $\dd \mu(V)$ is the Haar measure over the unitary group $U(k)$. Performing also the change of variables $M \to V M V^\dagger$ whose Jacobian is $1$ since $\det(V^\dagger V) = 1$, we find
\begin{equation}
  \moyst{\Tr[X_1]}
  = \frac{1}{2}
  \frac{
    \displaystyle
    \int_{\mathbb{R}_+^k} \dd^k h
    \: \Delta(h)^2
    \left( \sum_i h_i \right)
    \prod_{i} h_i^{N-k}
    \e^{- \frac{1}{T} \left( \frac{h_i^3}{12} + E h_i \right)}
    \int \dd M \:
    \e^{- \frac{1}{T} \sum_i h_i (M^2)_{ii}}
  }{
    \displaystyle
    \int_{\mathbb{R}_+^k} \dd^k h
    \: \Delta(h)^2
    \prod_{i} h_i^{N-k}
    \e^{- \frac{1}{T} \left( \frac{h_i^3}{12} + E h_i \right)}
    \int \dd M \:
    \e^{- \frac{1}{T} \sum_i h_i (M^2)_{ii}}
  }
  \:.
\end{equation}
There remains to evaluate the Gaussian integral over the matrix $M$. This can be done by writing it in terms of independent real components,
\begin{equation}
  \sum_i h_i (M^2)_{ii}
  = \sum_i h_i M_{ii}^2
  + \sum_{i<j} (h_i + h_j)(\mathrm{Re}[M_{ij}]^2 + \mathrm{Im}[M_{ij}]^2)
  \:.
\end{equation}
Hence,
\begin{align}
  \int \dd M \:
  \e^{- \frac{1}{T} \sum_i h_i (M^2)_{ii}}
  &= \left( \prod_i \int \dd M_{ii} \e^{- \frac{h_i}{T} M_{ii}^2} \right)
  \prod_{i<j}\left( \int \dd \mathrm{Re}[M_{ij}] \: \e^{-\frac{h_i + h_j}{T} \mathrm{Re}[M_{ij}]^2} \right)
    \left( \int \dd \mathrm{Im}[M_{ij}] \: \e^{-\frac{h_i + h_j}{T} \mathrm{Im}[M_{ij}]^2} \right)
    \nonumber
  \\
  & \propto \prod_i \frac{1}{\sqrt{h_i}}
    \prod_{i<j} \frac{1}{h_i+h_j}
    \:.
\end{align}
We thus obtain that
\begin{equation}
  \moyst{\Tr[X_1]}
  = \frac{1}{2}
  \frac{
    \displaystyle
    \int_{\mathbb{R}_+^k} \dd^k h
    \left( \sum_i h_i \right)
    \prod_{i<j} \frac{(h_j - h_i)^2}{h_i+h_j}
    \prod_{i} h_i^{N-k-\frac{1}{2}}
    \e^{- \frac{1}{T} \left( \frac{h_i^3}{12} + E h_i \right)}
  }{
    \displaystyle
    \int_{\mathbb{R}_+^k} \dd^k h
    \:
    \prod_{i<j} \frac{(h_j - h_i)^2}{h_i+h_j}
    \prod_{i} h_i^{N-k-\frac{1}{2}}
    \e^{- \frac{1}{T} \left( \frac{h_i^3}{12} + E h_i \right)}
  }
  \:.
\end{equation}
Defining
\begin{equation}
  \label{eq:DefZnkSM}
  \Z_{N,k}(E,T) = \int_{\mathbb{R}_+^k} \dd^k h
    \:
    \prod_{i<j} \frac{(h_j - h_i)^2}{h_i+h_j}
    \prod_{i} h_i^{N-k-\frac{1}{2}}
    \e^{- \frac{1}{T} \left( \frac{h_i^3}{12} + E h_i \right)}
    \:,
\end{equation}
we finally obtain
\begin{equation}
  \label{eq:ExprSumLyapsSM}
  \moyst{\Tr[X_1]} = \sum_{n=1}^k \gamma_n(E,T) = -\frac{T}{2} \partial_E \ln \Z_{N,k}(E,T)
  \:,
\end{equation}
corresponding to Eq.~(21) of the main text.

\section{Expression of the integrated density of states}

Fyodorov and Le Doussal have shown that, for $q>0$~\cite{Fyodorov:2026}
\begin{equation}
  \label{eq:GenFctTrZSM}
  \moyst{\e^{- \I q \Tr[Z]}}
  \underset{q \to 0}{\simeq}
  1 - q (\pi \nu_N(E,T) + \I \Gamma_N(E,T))
  \:,
\end{equation}
where $\nu_N(E,T)$ is the integrated density of states, $\Gamma_N(E,T)$ the sum of the Lyapunov exponents and $Z$ is the Riccati matrix that obeys
\begin{equation}
  \label{eq:RiccatiMatSM}
  \partial_x Z = -Z^2 - E \un_N + W
  \:.
\end{equation}
We show here that from the stationary distribution~\eqref{eq:JointStatioDistrSM} and the behavior~\eqref{eq:GenFctTrZSM}, we can recover the expression of the density of states given in~\cite{Fyodorov:2026b}, which is an alternative form to the one first derived in~\cite{Fyodorov:2026}.

For $k=N$, the block $X_2$ disappears, so that the equation for $X_1$~(\ref{eq:JointEqsMatX}) becomes identical to the equation satisfied by the Riccati matrix $Z$~(\ref{eq:RiccatiMatSM}). We can therefore compute the l.h.s. of~(\ref{eq:GenFctTrZSM}) using the stationary distribution~(\ref{eq:JointStatioDistrSM}),
\begin{equation}
  \moyst{\e^{- \I q \Tr[Z]}}
  = \frac{
    \displaystyle
    \int \dd X_1 \int_{H>0} \dd H \:
    \e^{- \frac{1}{T} \Tr[H X_1^2 - X_1 H^2 + \frac{H^3}{3} + E H] - \I q \Tr[X_1]}
  }{
    \displaystyle
    \int \dd X_1 \int_{H>0} \dd H \:
    \e^{- \frac{1}{T} \Tr[H X_1^2 - X_1 H^2 + \frac{H^3}{3} + E H]}
  }
  \:.
\end{equation}
The integrals over $X_1$ are Gausian, so they can be computed exactly. Explicitly, writing
\begin{equation}
  \frac{1}{T} \Tr[H X_1^2 - X_1 H^2] + \I q \Tr[X_1]
  = \frac{1}{T} \Tr \left[H \left(X_1 - \frac{1}{2} H + \frac{\I q T}{2} H^{-1} \right)^2 \right]
  - \frac{1}{4T} \Tr[H^3] + \frac{q^2 T}{4} \Tr[H^{-1}] + \frac{\I q T}{2} \Tr[H]
  \:,
\end{equation}
we obtain
\begin{equation}
  \moyst{\e^{- \I q \Tr[Z]}}
  = \frac{
    \displaystyle
    \int_{H>0} \dd H \:
    \e^{- \frac{1}{T} \Tr[\frac{H^3}{12} + E H] - \frac{q^2 T}{4} \Tr[H^{-1}] - \frac{\I q T}{2} \Tr[H]}
    \int \dd M \e^{- \frac{1}{T} \Tr[H M^2]}
  }{
    \displaystyle
    \int_{H>0} \dd H \:
    \e^{- \frac{1}{T} \Tr[\frac{H^3}{12} + E H]}
    \int \dd M \e^{- \frac{1}{T} \Tr[H M^2]}
  }
  \:,
\end{equation}
with $M$ an Hermitian matrix. Performing the same changes of variables as in Section~\ref{sec:Lyaps} above to compute the integrals over $M$, we obtain
\begin{equation}
   \moyst{\e^{- \I q \Tr[Z]}}
  =
  \frac{
    \displaystyle
    \int_{\mathbb{R}_+^N} \dd^N h
    \prod_{i<j} \frac{(h_j - h_i)^2}{h_i+h_j}
    \prod_{i} h_i^{-\frac{1}{2}}
    \e^{- \frac{1}{T} \left( \frac{h_i^3}{12} + E h_i \right)}
    \: \prod_{i} \e^{- \frac{\I q T}{2} h_i - \frac{q^2 T}{4 h_i}}
  }{
    \displaystyle
    \int_{\mathbb{R}_+^N} \dd^N h
    \:
    \prod_{i<j} \frac{(h_j - h_i)^2}{h_i+h_j}
    \prod_{i} h_i^{-\frac{1}{2}}
    \e^{- \frac{1}{T} \left( \frac{h_i^3}{12} + E h_i \right)}
  }
  \:.
\end{equation}
Equivalently,
\begin{equation}
  \label{eq:GenFctIntSM}
   \moyst{\e^{- \I q \Tr[Z]}}
   =
   1 - 
  \frac{
    \displaystyle
    \int_{\mathbb{R}_+^N} \dd^N h
    \prod_{i<j} \frac{(h_j - h_i)^2}{h_i+h_j}
    \prod_{i} h_i^{-\frac{1}{2}}
    \e^{- \frac{1}{T} \left( \frac{h_i^3}{12} + E h_i \right)}
    \left( 1 -  \prod_{i} \e^{- \frac{\I q T}{2} h_i - \frac{q^2 T}{4 h_i}} \right)
  }{
    \displaystyle
    \int_{\mathbb{R}_+^N} \dd^N h
    \:
    \prod_{i<j} \frac{(h_j - h_i)^2}{h_i+h_j}
    \prod_{i} h_i^{-\frac{1}{2}}
    \e^{- \frac{1}{T} \left( \frac{h_i^3}{12} + E h_i \right)}
  }
  \:.
\end{equation}
Expanding naively this expression at first order in $q$, we get only an imaginary term proportional to $q$. This is in contradiction with the form~(\ref{eq:GenFctTrZSM}). The reason can be seen by looking at the expansion of~(\ref{eq:GenFctIntSM}) at second order in $q$, which generates integrals which diverge at $h_i=0$. This shows that the small $q$ expansion of~(\ref{eq:GenFctIntSM}) is singular and must be treated carefully. The divergence at $0$ gives a clue on how to regularize the integral: it is caused by an eigenvalue of $H$ becoming too small. For $q \neq 0$ this is normally compensated by the term $\e^{- \frac{q^2 T}{4h_i}}$ which appears subleading at linear order in $q$. We thus choose one of $N$ the eigenvalues, say $h_1$, to be the smallest and make the change of variables $h_1 = \frac{q^2 T}{4x}$. We get,
\begin{multline}
   1 - \mathrm{Re}\left[ \moyst{\e^{- \I q \Tr[Z]}} \right]
   =
   \frac{N}{\Z_{N,N}(E,T)} \frac{|q| \sqrt{T}}{2}
    \int_0^\infty \frac{\dd x}{x^{3/2}}
    \int_{\frac{q^2 T}{4x}}^\infty \dd^{N-1} h
    \prod_{2<i<j} \frac{(h_j - h_i)^2}{h_i+h_j}
    \\
    \times
    \prod_{i=2}^{N} h_i^{-\frac{1}{2}}
    \e^{- \frac{1}{T} \left( \frac{h_i^3}{12} + E h_i \right)}
    \frac{\left( \frac{q^2 T}{4 x} - h_i \right)^2}{\frac{q^2 T}{4 x} - h_i}
    \left( 1 - \e^{- \frac{\I q^3 T^2}{8 x} - x} \prod_{i=2}^{N} \e^{- \frac{\I q T}{2} h_i - \frac{q^2 T}{4 h_i}} \right)
  \:.
\end{multline}
For $q \to 0$, this gives
\begin{multline}
   1 - \mathrm{Re}\left[ \moyst{\e^{- \I q \Tr[Z]}} \right]
   =
   \frac{N}{\Z_{N,N}(E,T)} \frac{|q| \sqrt{T}}{2}
    \int_0^\infty \frac{\dd x}{x^{3/2}}
    \int_{0}^\infty \dd^{N-1} h
    \prod_{2<i<j} \frac{(h_j - h_i)^2}{h_i+h_j}
    \\
    \times
    \prod_{i=2}^{N} h_i^{1-\frac{1}{2}}
    \e^{- \frac{1}{T} \left( \frac{h_i^3}{12} + E h_i \right)}
    \left( 1 - \e^{- x} \right)
  \:.
\end{multline}
Since $\int_0^\infty \frac{\dd x}{x^{3/2}}(1-\e^{-x}) = 2 \sqrt{\pi}$,
we finally obtain
\begin{equation}
\label{eq:BehavSmallq}
  1 - \mathrm{Re}\left[ \moyst{\e^{- \I q \Tr[Z]}} \right]
  = |q| N \sqrt{\pi T} \: \frac{\Z_{N,N-1}(E,T)}{\Z_{N,N}(E,T)}
  \:,
\end{equation}
where we have recognized in the numerator the integral over the $N-1$ variables $h_i$ to be $\Z_{N,N-1}$. This is Eq.~(23) of the main text, from which the expression of the density of states follows
from~(\ref{eq:GenFctTrZSM}),
\begin{equation}
  \label{eq:IDoSSM}
  \nu_N(E,T) = N \sqrt{\frac{T}{\pi}} \frac{\Z_{N,N-1}(E,T)}{\Z_{N,N}(E,T)}
  \:,
\end{equation}
recovering the result of~\cite{Fyodorov:2026b} and corresponding to Eq.~(24) of the main text.

\section{Asymptotic analysis of $\Z_{N,k}$ for large energy}
\label{sec:LargeE}

Let us first consider the limit $E \to \infty$. We make the change of variables $h_i \to h_i/E$ in the expression~(\ref{eq:DefZnkSM}) of $\Z_{N,k}$,
\begin{equation}
  \Z_{N,k} = E^{-N k + \frac{k^2}{2}}
  \int_0^\infty \dd^k h \prod_{i<j} \frac{(h_j-h_i)^2}{h_i+h_j}
  \prod_i h_i^{N-k-\frac{1}{2}}
  \e^{- \frac{1}{T} \left(\frac{h_i^3}{12 E^3} + h_i \right)}
  \:.
\end{equation}
The multiple integral converges for $E \to \infty$ and becomes independent of $E$, thus
\begin{equation}
  \Z_{N,k}(E,T) \underset{E \to \infty}{\propto}
  E^{-N k + \frac{k^2}{2}}
  \:,
\end{equation}
or equivalently,
\begin{equation}
  \partial_E \ln \Z_{N,k}(E,T) \underset{E \to \infty}{\simeq}
  \frac{1}{E} \left( \frac{k^2}{2} - N k  \right)
  \:.
\end{equation}
Using this expression in~(\ref{eq:ExprSumLyapsSM}) and taking the difference between $k$ and $k-1$ yields
\begin{equation}
  \gamma_k(E,T) = \frac{T}{2E} \left(
    N - k + \frac{1}{2}
  \right)
  \:,
\end{equation}
which corresponds to Eq.~(26) of the main text.

\bigskip

The analysis of the limit $E \to -\infty$ is a bit more involved, since it requires a saddle point estimate. Explicitly, we make the change of variables
\begin{equation}
  h_i = 2 \sqrt{-E} + \frac{x_i}{(-E)^{\frac{1}{4}}}
  \:.
\end{equation}
At leading order for $E \to -\infty$, we get Eq.~(\ref{eq:DefZnkSM})
becomes
\begin{equation}
  \Z_{N,k} \underset{E \to -\infty}{\simeq}
  2^{\frac{k}{2}(2N-4k+1)} (-E)^{\frac{Nk}{2} - k^2} \e^{\frac{4k}{3T}(-E)^{\frac{3}{2}}}
  \int \dd^k x \: \prod_{i<j} (x_j-x_i)^2 \prod_i \e^{-\frac{x_i^2}{2T}}
  \:.
\end{equation}
The last integral does not depend on $E$, hence,
\begin{equation}
  \partial_E \ln \Z_{N,k}(E,T)
  \underset{E \to -\infty}{\simeq}
  - \frac{2k}{T} \sqrt{-E} + \frac{1}{2E} \left(
    N k - 2 k^2
  \right)
  \:.
\end{equation}
Using again the relation with the Lyapunov
exponents~(\ref{eq:ExprSumLyapsSM}), we obtain
\begin{equation}
  \gamma_k(E,T) \underset{E \to -\infty}{\simeq}
  \sqrt{-E}
  + \frac{T}{4(-E)}(N - 4 k + 2)
  \:,
\end{equation}
corresponding to Eq.~(25) of the main text.

\section{Large $N$ analysis of the smallest Lyapunov exponent}

Combining the expression of the Lyapunov exponents~(\ref{eq:ExprSumLyapsSM}) with the integrated density of states~(\ref{eq:IDoSSM}), we can write the smallest Lyapunov exponent as
\begin{equation}
  \label{eq:LyapIDoSSM}
  \gamma_N(E,T) = \frac{T}{2} \partial_E \ln \nu_N(E,T)
  \:.
\end{equation}
This allows us to study the behavior of the smallest Lyapunov exponent by relying on the results obtained previously on the integrated density of states~\cite{Allez:2015,Fyodorov:2020,Fyodorov:2026a,Fyodorov:2026}.

In particular, for $NT = J^2/2$ fixed, we obtain from~(S115) of Ref.~\cite{Fyodorov:2026} and the following equations,
\begin{equation}
  \label{eq:IDoSPos}
  \lim_{N \to \infty} \frac{\nu_N(E)}{N}
  = - \frac{1}{\pi} \mathrm{Im} \left(
    d(E) + \frac{J^2}{8 d(E)}
  \right)
  \equiv n(E)
  \:,
\end{equation}
where $d(E)$ is the solution of
\begin{equation}
  \label{eq:CubicEq}
  d^3 + E d + \frac{J^2}{2} = 0
  \:,
\end{equation}
which behaves as $-\I \sqrt{E}$ for large positive $E$. We denote $E_\star = -3s^2/4$ with $s = (2J^2)^{1/3}$. For $E < E_\star$, the three solutions of~(\ref{eq:CubicEq}) are real, so the integrated
density of states per channel~(\ref{eq:IDoSPos}) vanishes.  For $E > E_\star$, applying relation~(\ref{eq:LyapIDoSSM}) we find 
\begin{equation}
  \gamma_N(E,T) \underset{N \to \infty}{\simeq} \frac{J^2}{4N} \partial_E n(E)
  \:,
\end{equation}
which tends to zero as $N \to \infty$.

\medskip

For $E < E_\star$, we use the analysis presented in the Appendix C of Ref.~\cite{Fyodorov:2026a} which gives a large deviation form for the density of states,
\begin{equation}
  \ln \nu_N(E,T)
  \underset{N \to \infty}{\simeq}
  - \frac{2N}{J^2} U(E)
  \:,
  \quad
  U(E) = \int_{-d(E)}^{\gamma_-(E)}
  (\lambda + d(E)) \sqrt{(\gamma_-(E)-\lambda)(\gamma_+(E)-\lambda)} \dd \lambda
  \:,
\end{equation}
where $d(E)$ is now the largest real root of~(\ref{eq:CubicEq}) and
\begin{equation}
  \gamma_\pm(E) = d(E) \pm \frac{J}{\sqrt{d(E)}}
  \:.
\end{equation}
Using this form into the relation~(\ref{eq:LyapIDoSSM}) yields
\begin{equation}
  \gamma_N(E,T) \underset{N \to \infty}{\simeq}
  \sqrt{d(E)^2 - \frac{J^2}{4 d(E)}}
  \:.
\end{equation}

\section{Determinantal form of the multiple integrals}

We have obtained the Lyapunov exponents in terms of the $k$-dimensional integrals $\Z_{N,k}$~(\ref{eq:DefZnkSM}). Although these expressions are well suited for analytical calculations (see Section~\ref{sec:LargeE}), it is not so convenient for numerical calculations. The goal of this Section is to rewrite~(\ref{eq:DefZnkSM}) as a determinant, which can be computed numerically more easily.

For this, we follow the classical approach of De Bruijn and Andreief~\cite{DeBruijn:1955,Andreief:1986}. The first step is to expand one Vandermonde determinant as
\begin{equation}
  \Delta(h) = \det[h_i^{j-1}]_{1 \leq i,j \leq k}
  = \sum_{\sigma \in S_k} \sg{\sigma}
  \prod_{i=1}^k h_i^{\sigma(i)-1}
  \:,
\end{equation}
where $S_k$ is the ensemble of permutations of $k$ integers. Inserting this expression into the definition of $\Z_{N,k}$~(\ref{eq:DefZnkSM}), we obtain
\begin{equation}
  \Z_{N,k}(E,T) = \sum_{\sigma \in S_k} \sg{\sigma}
  \int_{\mathbb{R}_+^k} \dd^k h
    \:
    \prod_{i<j} \frac{(h_j - h_i)}{h_i+h_j}
    \prod_{i} h_i^{N-k-\frac{3}{2} + \sigma(i)}
    \e^{- \frac{1}{T} \left( \frac{h_i^3}{12} + E h_i \right)}
    \:.
\end{equation}
Relabelling the indices of the integration variables with $\sigma$ and using that $\prod_{i<j}(h_{\sigma(i)} - h_{\sigma(j)}) = \sg{\sigma} \prod_{i<j}(h_{i} - h_{j)})$, we obtain
\begin{equation}
  \label{eq:ExprZinterm}
  \Z_{N,k}(E,T) = k!
  \int_{\mathbb{R}_+^k} \dd^k h
    \:
    \prod_{i<j} \frac{(h_j - h_i)}{h_i+h_j}
    \prod_{i} h_i^{N-k-\frac{3}{2} + i}
    \e^{- \frac{1}{T} \left( \frac{h_i^3}{12} + E h_i \right)}
    \:.
\end{equation}
For the remaining terms, we use Schur's identity~\cite{Schur:1911},
\begin{equation}
  \label{eq:Schur}
  \prod_{i<j}^{2n} \frac{(h_j - h_i)}{h_i+h_j}
  = \mathrm{Pf} \left[ \frac{h_j - h_i}{h_j + h_i} \right]_{1 \leq i,j \leq 2n}
  \:,
  \quad
  \prod_{i<j}^{2n+1} \frac{(h_j - h_i)}{h_i+h_j}
  = \mathrm{Pf} \left[
    \begin{pmatrix}
      \left( \frac{h_j - h_i}{h_j + h_i} \right)_{1 \leq i,j \leq 2n} & 1
      \\
      -1 & 0
    \end{pmatrix}
    \right]
    \:,
\end{equation}
where $\mathrm{Pf}$ is a Pfaffian. We thus need to distinguish the different parities of $k$.

\medskip

For $k = 2n$ even, we write
\begin{equation}
   \prod_{i<j}^{2n} \frac{(h_j - h_i)}{h_i+h_j}
   = \mathrm{Pf} \left[ \frac{h_j - h_i}{h_j + h_i} \right]_{1 \leq i,j \leq 2n}
   = \frac{1}{2^n n!} \sum_{\sigma \in S_{2n}} \sg{\sigma}
   \prod_{i=1}^n \frac{h_{\sigma(2i-1)} - h_{\sigma(2i)}}{h_{\sigma(2i-1)} + h_{\sigma(2i)}}
   \:.
\end{equation}
Using this expression into the multiple integral~(\ref{eq:ExprZinterm}) gives
\begin{align}
  \label{eq:ZnkPfEven}
  \Z_{N,k}
  &= \frac{k!}{2^n n!}
  \sum_{\sigma \in S_{2n}} \sg{\sigma}
    \int_0^\infty \dd^k h \prod_{i=1}^n \frac{h_{\sigma(2i-1)} - h_{\sigma(2i)}}{h_{\sigma(2i-1)} + h_{\sigma(2i)}}
    h_{\sigma(2i-1)}^{\sigma(2i-1)-1} w_{N-k}(h_{\sigma(2i-1)})
  h_{\sigma(2i)}^{\sigma(2i)-1} w_{N-k}(h_{\sigma(2i)})
    \nonumber
  \\
  &=  \frac{k!}{2^n n!}
    \sum_{\sigma \in S_{2n}} \sg{\sigma}
    \prod_{i=1}^n \int_0^\infty \dd h_1 \int_0^\infty \dd h_2 \frac{h_2 - h_1}{h_1+h_2}
    h_1^{\sigma(2i-1)-1} h_2^{\sigma(2i)-1}
    w(h_1) w(h_2)
    \nonumber
  \\
  &=
    k! \: \mathrm{Pf} \left[
    \int_0^\infty \dd h_1 \int_0^\infty \dd h_2 \frac{h_2 - h_1}{h_1+h_2}
    h_1^{a-1} h_2^{b-1}
    w_{N-k}(h_1) w_{N-k}(h_2)
    \right]_{1 \leq a,b \leq 2n}
    \:,
    \nonumber
  \\
  &= k! \: \mathrm{Pf} \left[
    M_{a,b}
    \right]_{1 \leq a,b \leq 2n}
    \:,
\end{align}
where
\begin{equation}
  w_p(h) = h^{p-\frac{1}{2}} \e^{- \frac{1}{T}(\frac{h^3}{12} + E h)}
\end{equation}
and
\begin{equation}
  \label{eq:DefMatM}
  M_{a,b} =     \int_0^\infty \dd h_1 \int_0^\infty \dd h_2 \frac{h_2 - h_1}{h_1+h_2}
  h_1^{a-1} h_2^{b-1}
  w_{N-k}(h_1) w_{N-k}(h_2)
  \:.
\end{equation}

\medskip

Performing the same analysis for $k=2n+1$ using~(\ref{eq:Schur}), we find
\begin{equation}
  \label{eq:ZnkPfOdd}
  \Z_{N,k} = k! \: \mathrm{Pf} \left[
    \begin{pmatrix}
      M & m
      \\
      - m^{\mathrm{T}} & 0
    \end{pmatrix}
  \right]
  \:,
\end{equation}
where $M$ is given by~(\ref{eq:DefMatM}) and is of size $k \times k$ and $m$ is the column vector
\begin{equation}
  m_a = \int_0^\infty h^{a-1} w_{N-k}(h) \dd h
  \:.
\end{equation}

\medskip

The two expressions~(\ref{eq:ZnkPfEven},\ref{eq:ZnkPfOdd}) can actually be combined into a single formula. For this, we introduce the $k \times k$ matrix of elements
\begin{equation}
  \label{eq:MatC}
  C_{ab} =  \int_0^\infty \dd h_1 \int_0^\infty \dd h_2 \frac{h_1^{a-1} h_2^{b}}{h_1+h_2}
  w_{N-k}(h_1) w_{N-k}(h_2)
  \:.
\end{equation}
From this definition, we have
\begin{equation}
  C_{ab} + C_{ba} = m_a m_b
  \:,
  \quad
  C_{ab} - C_{ba} = M_{ab}
  \:.
\end{equation}
Therefore, we can write
\begin{equation}
  2 C_{ab} = M_{ab} + m_a m_b
  \:.
\end{equation}
Taking the determinant of this expression and using the properties of the determinant of a rank $1$ perturbation, we get for even $k$ (so that $M$ is invertible)
\begin{equation}
  \det(2C) = \det(M) (1 + m^{\mathrm{T}} M^{-1} m)
  = \det(M)
  \:,
\end{equation}
since $M$ is antisymmetric, $m^{\mathrm{T}} M^{-1} m = (m^{\mathrm{T}} M^{-1} m)^{\mathrm{T}} = -
m^{\mathrm{T}} M^{-1} m = 0$. Since $\det(M) = \mathrm{Pf}[M]^2$ and $\Z_{N,k} > 0$, we get
\begin{equation}
  \label{eq:ZnkEvenWithC}
  \Z_{N,k} = k! 2^{\frac{k}{2}} \sqrt{\det(C)}
  \:, \quad \text{for } k \text{ even.}
\end{equation}
For odd $k$ we compute the determinant in~(\ref{eq:ZnkPfOdd}) by expanding the last row and column. This gives
\begin{equation}
   \det \left[
    \begin{pmatrix}
      M & m
      \\
      - m^{\mathrm{T}} & 0
    \end{pmatrix}
  \right]
  = \sum_{i,j=1}^{k} (-1)^{i+j} m_i m_j \det[M_{(i,j)}]
  = m^{\mathrm{T}} \mathrm{adj}(M) m
  \:,
\end{equation}
where $M_{(i,j)}$ corresponds to matrix $M$ with row $i$ and column $j$ removed and $\mathrm{adj}(M)$ is the adjugate matrix of $M$ (which is not invertible).  Using the matrix determinant lemma $\det(M + m m^{\mathrm{T}}) = \det(M) + m^{\mathrm{T}} \mathrm{adj}(M) m $, we thus obtain
\begin{equation}
   \det \left[
    \begin{pmatrix}
      M & m
      \\
      - m^{\mathrm{T}} & 0
    \end{pmatrix}
  \right]
  = \det(M + m m^{\mathrm{T}})
  = \det(2C)
  \:.
\end{equation}
Thus, we obtain the same formula as~(\ref{eq:ZnkEvenWithC}), but for odd $k$. Therefore, for any $k$,
\begin{equation}
  \label{eq:ZnkAllWithC}
  \Z_{N,k} = k! 2^{\frac{k}{2}} \sqrt{\det(C)}
  \:.
\end{equation}
This gives a practical way to compute numerically $\Z_{N,k}$ from the matrix $C$~(\ref{eq:MatC}).

\bigskip

The Lyapunov exponents are obtained from derivatives with respect to $E$ of $\Z_{N,k}$ via relation~(\ref{eq:ExprSumLyapsSM}). From the Jacobi relation, we can write the derivative of~(\ref{eq:ZnkAllWithC}) as
\begin{equation}
  \partial_E \ln \Z_{N,k} = \frac{1}{2} \Tr[C^{-1} \partial_E C]
  \:.
\end{equation}
Using the expression of $C$~(\ref{eq:MatC}), the derivative with respect to $E$ simplifies as
\begin{equation}
  \partial_E C_{ab} = - \frac{1}{T}
   \int_0^\infty \dd h_1 \int_0^\infty \dd h_2 \:  h_1^{a-1} h_2^{b}
  w_{N-k}(h_1) w_{N-k}(h_2)
  = - \frac{1}{T} m_{a} m_{b+1}
  \:.
\end{equation}
Hence, we obtain
\begin{equation}
  \partial_E \ln \Z_{N,k}
  = - \frac{1}{2T} m_+^{\mathrm{T}} C^{-1} m
  \:,
  \quad
  \text{with}
  \quad
  (m_+)_a = m_{a+1}
  \:.
\end{equation}
This expression gives a practical way to compute the Lyapunov exponents as
\begin{equation}
  \moyst{\Tr[X_1]} = \sum_{n=1}^k \gamma_n(E,T)
  = \frac{1}{4} m_+^{\mathrm{T}} \tilde{m}
  \:,
  \quad \text{with } \tilde{m} \text{ solution of }
  m = C \tilde{m}
  \:.
\end{equation}

\section{Numerical simulations}

To compute numerically the Lyapunov exponents, it is convenient to rewrite the multichannel Schrödinger equation $- \partial_x^2 \Psi + W \Psi = E \Psi$ as the first order system
\begin{equation}
  \label{eq:MatSchrod}
  \partial_x
  \begin{pmatrix}
    \partial_x \Psi
    \\
    \Psi
  \end{pmatrix}
  =
  \begin{pmatrix}
    0 & W-E
    \\
    \un_N  & 0
  \end{pmatrix}
  \begin{pmatrix}
    \partial_x \Psi
    \\
    \Psi
  \end{pmatrix}
  \:.
\end{equation}
The goal is now to discretize this expression and express the vector $(\partial_x \Psi \: \Psi)^{\mathrm{T}}$ as a product of matrices. It is convenient to write
\begin{equation}
  \label{eq:PotDirac}
  W(x) = \sum_n W_n \: \delta(x-x_n)
  \:,
  \quad
  x_n = n \delta x
  \:,
\end{equation}
with $W_n$ are independent and identically distributed from
\begin{equation}
  P[W_n] \propto \exp \left[
    - \frac{1}{4T \delta x} \Tr[W_n^\dagger W_n]
  \right]
  \:.
\end{equation}
When $\delta x \to 0$, this reduces to the white noise potential considered here~\cite{Texier:1999}. With this form of the noise, Eq.~(\ref{eq:MatSchrod}) becomes a free evolution on all intervals
$]x_n,x_{n+1}[$, with a jump at $x_n$ given by $W_n$. We can thus write
\begin{equation}
  \begin{pmatrix}
    \partial_x \Psi
    \\
    \Psi
  \end{pmatrix}(x_{n}^+)
  =
  \begin{pmatrix}
    \un_N & W_{n}
    \\
    0 & \un_N
  \end{pmatrix}
  \begin{pmatrix}
    \cos(k \delta x) & - k \sin(k \delta x)
    \\
    \frac{\sin(k \delta x)}{k} & \cos(k \delta x)
  \end{pmatrix}
  \begin{pmatrix}
    \partial_x \Psi
    \\
    \Psi
  \end{pmatrix}(x_{n-1}^+)
  \:,
  \quad
  k = \sqrt{E}
  \;,
\end{equation}
for $E>0$. In the case $E<0$, we have
\begin{equation}
  \begin{pmatrix}
    \partial_x \Psi
    \\
    \Psi
  \end{pmatrix}(x_{n}^+)
  =
  \begin{pmatrix}
    \un_N & W_{n}
    \\
    0 & \un_N
  \end{pmatrix}
  \begin{pmatrix}
    \cosh(k \delta x) &  k \sinh(k \delta x)
    \\
    \frac{\sinh(k \delta x)}{k} & \cosh(k \delta x)
  \end{pmatrix}
  \begin{pmatrix}
    \partial_x \Psi
    \\
    \Psi
  \end{pmatrix}(x_{n-1}^+)
  \:,
  \quad
  k = \sqrt{-E}
  \;.
\end{equation}
In both cases, we can write
\begin{equation}
  \label{eq:ProdMatrices}
  \begin{pmatrix}
    \partial_x \Psi
    \\
    \Psi
  \end{pmatrix}(x_{n}^+)
  = A_n A_{n-1} \cdots A_1
  \begin{pmatrix}
    \partial_x \Psi
    \\
    \Psi
  \end{pmatrix}(0^+)
  \:,
\end{equation}
with
\begin{equation}
  A_n = \left\lbrace
    \begin{array}{ll}
      \displaystyle
      \begin{pmatrix}
        \un_N & W_{n}
        \\
        0 & \un_N
      \end{pmatrix}
      \begin{pmatrix}
        \cosh(k \delta x) &  k \sinh(k \delta x)
        \\
        \frac{\sinh(k \delta x)}{k} & \cosh(k \delta x)
      \end{pmatrix}
      & \text{for } E = -k^2
        \:,
      \\[0.5cm]
      \displaystyle
      \begin{pmatrix}
        \un_N & W_{n}
        \\
        0 & \un_N
      \end{pmatrix}
      \begin{pmatrix}
        \cos(k \delta x) & - k \sin(k \delta x)
        \\
        \frac{\sin(k \delta x)}{k} & \cos(k \delta x)
      \end{pmatrix}
      & \text{for } E = k^2
        \:.
    \end{array}
  \right.
\end{equation}

The Lyapunov exponents associated with the product of matrices~(\ref{eq:ProdMatrices}) can be computed using standard techniques~\cite{Dieci:1995,Dieci:2010} based on the QR decomposition. The method is as follows. For any $n$, we introduce the QR decomposition of the product
\begin{equation}
  \Pi_n \equiv A_n A_{n-1} \cdots A_1 = Q_n R_n
  \:,
\end{equation}
where $Q_n$ is unitary and $R_n$ upper triangular with positive diagonal elements. By definition, we have
\begin{equation}
  \Pi_{n+1} = A_{n+1} \Pi_{n} = A_{n+1} Q_n R_n
  \:.
\end{equation}
Denoting
\begin{equation}
  \tilde{Q}_n \tilde{R}_n = A_{n+1} Q_n
\end{equation}
the $QR$ decomposition of $A_{n+1}Q_n$, we can write recursion relations for $Q_n$ and $R_n$ using the uniqueness of the QR decomposition,
\begin{equation}
  Q_{n+1} = \tilde{Q}_n
  \:,
  \quad
  R_{n+1} = \tilde{R}_n R_n
  \:.
\end{equation}
As discussed in the main text, the Lyapunov exponents can be extracted from the diagonal elements of $R_n$~\cite{Dieci:1995,Dieci:2010},
\begin{equation}
  \label{eq:LyapsQRSM}
  \gamma_k = \lim_{n \to \infty} \frac{1}{n} \ln[(R_n)_{kk}]
  =  \lim_{n \to \infty} \frac{1}{n} \sum_{k=1}^n \ln[(\tilde{R}_n)_{kk}]
  \:.
\end{equation}
This provides an algorithm to compute the Lyapunov exponents numerically: starting from $Q_0 = \un_{2N}$, we iterate the QR decompositions of $A_{n+1}Q_n = \tilde{Q}_n \tilde{R}_n$. We then set $Q_{n+1} = \tilde{Q}_n$ and store the diagonal elements to compute~(\ref{eq:LyapsQRSM}).

In practice, we evaluate~(\ref{eq:LyapsQRSM}) for $n = 10^7$ steps, with $\delta x = 10^{-2}$.


%